\documentclass[10pt]{article}
\usepackage{mathptmx}
\usepackage{amssymb,amsmath}
\newtheorem{thm}{Theorem}[section]
\newtheorem{cor}[thm]{Corollary}
\newtheorem{lem}[thm]{Lemma}
\newtheorem{defn}[thm]{Definition}
\newtheorem{prop}[thm]{Proposition}
\def\proof{{\bf Proof. }}
\newtheorem{remark}[thm]{Remark}
\def\be{\begin{equation}}
\def\ee{\end{equation}}
\def\bea{\begin{eqnarray}}
\def\eea{\end{eqnarray}}
\def\bean{\begin{eqnarray*}}
\def\eean{\end{eqnarray*}}
\def\ea{\end{array}}
\def\ds{\displaystyle}
\def\nm{\noalign{\medskip}}
\def\G{{\mathcal{G}}}

\def\O{{\mathcal{O}}}
\def\W{{\mathcal{W}}}
\def\V{{\mathcal{V}}}
\def\E{{\mathcal{E}}}
\def\F{{\mathcal{F}}}
\def\P{{\mathcal{P}}}
\def\L{{\mathcal{L}}}
\def\H{{\mathcal{H}}}
\def\S{{\mathcal{S}}}
\def\D{{\mathcal{D}}}
\def\N{{\bf N}}
\newcommand{\field}[1]{\mathbb{#1}}
\newcommand{\rz}{\field{R}}
\newcommand{\cz}{\field{C}}
\newcommand{\nz}{\field{N}}
\def\11{{\rm 1~\hspace{-1.2ex}l} }
\def\d{{\rm{d}}}
\def\ra{{\rangle}}
\def\la{{\langle}}
\def\cqfd{{$\blacksquare$}}

\def\hbarr{{\varepsilon}}
\begin{document}
\title{Propagation of chaos for many-boson systems in one dimension
with a point pair-interaction}
\author{
Z.~Ammari\thanks{zied.ammari@univ-rennes1.fr} \hspace{.3in}
S.~Breteaux\thanks{sebastien.breteaux@univ-rennes1.fr}\\
\\ {\small IRMAR, Universit\'e de Rennes I, }\\ {\small UMR-CNRS 6625,
campus de Beaulieu, 35042 Rennes Cedex, France.
}}
\maketitle
\begin{abstract}
We consider the semiclassical limit of nonrelativistic quantum many-boson systems
with delta potential in one dimensional space. We prove that time evolved
coherent states behave semiclassically  as  squeezed states by a Bogoliubov time-dependent
affine transformation. This allows us to obtain properties analogous to those proved by Hepp
and Ginibre-Velo (\cite{Hep}, \cite{GiVe1,GiVe2}) and also to show propagation of chaos for Schr\"odinger
dynamics in the mean field limit. Thus, we provide a derivation of the cubic NLS equation in one dimension.
\end{abstract}

\section{Introduction}
The justification of the chaos conservation  hypothesis in quantum  many-body theory is the main concern
of the present paper. This well-know hypothesis finds its roots in statistical physics of
classical many-particle systems as a quantum counterpart. See, for instance \cite{MaSh},
\cite{Got} and references therein.

Non-relativistic  quantum systems of $\N$ bosons moving in $d$-dimensional space are commonly
described by the Schr\"odinger Hamiltonian
\bea
\label{int-ham}
{\rm H}_\N:=\sum_{i=1}^\N -\Delta_{x_i}+\sum_{i<j} V_\N(x_i-x_j)\,, \quad x\in\rz^{d}\,,
\eea
acting on  the space  of symmetric square-integrable functions $ L^{2}_s(\rz^{d \N})$ over $\rz^{d\N}$. Here $V_\N$
stands for an even real  pair-interaction potential.
The Hamiltonian (\ref{int-ham}), under appropriate conditions on $V_\N$, defines a self-adjoint operator
and hence the Schr\"odinger equation
\bea
\label{schrodinger}
i\partial_t \Psi_\N^t={\rm H}_\N \Psi_\N^t,
\eea
admits a unique solution for any initial data $\Psi_\N^0\in L^2(\rz^{d \N})$. The
interacting $\N$-boson dynamics (\ref{schrodinger}) are considered in the mean field scaling,
namely, when $\N$ is large and the pair-potential
is given by
$$
V_\N(x)=\frac{1}{\N} V(x)\,,
$$
with $V$ independent of $\N$. The chaos conservation hypothesis for the $\N$-boson system
(\ref{schrodinger}) amounts to the study of the asymptotics
of the $k$-particle correlation functions $\gamma_{k,\N}^t$ given by
\bea
\label{corldef}
\gamma_{k,\N}^t(x_1,\cdots,x_k;y_1,\cdots,y_k)=
\int_{\rz^{d(\N-k)}} \gamma_N^t(x_1,\cdots,x_{k},z_{k+1},\cdots,z_\N;y_1,\cdots,y_{k},z_{k+1}, z_\N) dz_{k+1}\cdots dz_{\N}\,,
\eea
where $\gamma_N^t=\Psi_\N^t(x_1,\cdots,x_\N) \overline{\Psi_\N^t(y_1,\cdots,y_\N)}$.
More precisely, this hypothesis holds if for an initial datum which factorizes  as
$$
\Psi_\N^0=\varphi_0(x_1)\cdots \varphi_0(x_\N)\, \quad\mbox{ such that }\;||\varphi_0||_{L^2(\rz^d)}=1\,,
$$
the $k$-particle correlation functions converges in the trace norm
\bea
\label{partcor}
\gamma_{k,\N}^t \ds  \stackrel{\N\rightarrow\infty}{\longrightarrow}  \varphi_t(x_1)\cdots\varphi_t(x_k)\,
\overline{\varphi_t(y_1)\cdots\varphi_t(y_k)},
\eea
where $\varphi_t$ solves the nonlinear Hartree equation
\bea
\label{hartree}
\left\{
\begin{array}{l}
i \partial_t \varphi= -\Delta \varphi+ V*|\varphi|^2 \varphi\\
\varphi_{|t=0}=\varphi_0\,.
\end{array}\right.
\eea
The  convergence of correlation functions  (\ref{partcor}) for the Schr\"odinger dynamics (\ref{schrodinger}) is equivalent to the statement below :
\bea
\label{correl}
\lim_{\N\to\infty}\la \Psi_\N^t, \O_\N \Psi_\N^t\ra&= &
\lim_{\N\to\infty} \int_{\rz^{2d k}} \gamma_{k,\N}^t(x_1,\cdots,x_k;y_1,\cdots,y_k) \tilde \O(y_1,\cdots,y_k;x_1,\cdots,x_k)  dx_1\cdots dx_kdy_1
\cdots dy_k \nonumber
\\&=&\la \varphi_t^{\otimes k}, \O \varphi_t^{\otimes k}\ra\,,
\eea
where $\O_\N$ are observables given by
$
\O_\N:=\O\otimes 1^{(\N-k)}$  acting on $L^{2}(\rz^{d\N})$ with $\O:L^{2}(\rz^{dk})\to L^{2}(\rz^{dk})$
a bounded operator with kernel $\tilde\O$ and $k$ is a fixed integer. The relevance of those observables is justified by the fact that
$\O_\N$ are essentially canonical quantizations of classical quantities.

In the recent years, mainly motivated by the study of Bose-Einstein condensates, there is a renewed and growing interest
in the analysis of many-body quantum dynamics in the mean field limit
(for instance see \cite{ABGT},\cite{BEGMY},\cite{BGM},\cite{ESY},\cite{EY},\cite{FGS},\cite{FKP},\cite{FKS}, etc.).
For a general presentation on the subject we refer the reader to the reviews \cite{Spo} and \cite{Gol}. Various
strategies were developed in order to prove the chaos conservation hypothesis or even stronger statements. One of the oldest
approaches is the so-called BBGKY hierarchy (named after Bogoliubov, Born, Green, Kirkwood, and Yvon) which consists in
considering the Heisenberg equation,
\bea
\label{heisenberg}
\left\{
\begin{array}{l}
\partial_t \rho_t= i [\rho_t, {\rm H}_\N],\\
\rho_{|t=0}=|\varphi_0^{\otimes \N}\ra\la\varphi_0^{\otimes \N}|\,,
\end{array}\right.
\eea
together with the finite chain of equations arising from (\ref{heisenberg}) by taking partial traces on $0\leq k\leq\N$ variables.
Since $\rho_t$ are trace class operators one can write the corresponding hierarchy of equations on the $k$-particle
correlation functions $\gamma_{k,\N}^t$:
\bean
\left\{
\begin{array}{lll}
i\partial_t \gamma_{k,\N}^t&=& \ds\sum_{i=1}^\N \left[-\Delta_{x_i}+\Delta_{y_i}\right]\gamma_{k,\N}^t+
\ds\frac{1}{\N}\sum_{1\leq i<j\leq k} \left[V(x_i-x_j)-V(y_i-y_j)\right] \gamma_{k,\N}^t \\ \nm\ds
&&+ \ds\frac{1}{\N}\sum_{1\leq i\leq k, k+1\leq j\leq \N} \int_{\rz^{(\N-k)d}}\left[V(x_i-x_j)-V(y_i-y_j)\right] \gamma_{\N}^t \;dx_{k+1}\cdots
dx_{\N}\\ \nm\ds
&&+\ds\frac{1}{\N}\sum_{k+1\leq i<j\leq \N} \int_{\rz^{(\N-k)d}}\left[V(x_i-x_j)-V(y_i-y_j)\right] \gamma_{\N}^t \;dx_{k+1}\cdots
dx_{\N}\\ \nm\ds
\gamma_{k,\N}^0&=&\varphi_0(x_1)\cdots\varphi_0(x_k)\overline{\varphi_0(y_1)\cdots \varphi_0(y_k)}\,.
\end{array}
\right.
\eean

An alternative approach to the chaos conservation hypothesis uses the second quantization framework
(details on this notions are recalled in Section \ref{prem}). Consider the Hamiltonian,
\begin{eqnarray*}
 && \varepsilon^{-1}H_{\varepsilon}=\int_{\rz^{d}}
\nabla a^{*}(x)\nabla
a(x)~dx+\frac{\varepsilon}{2} \int_{\rz^{2d}}V(x-y)a^{*}(x)a^{*}(y)a(x)a(y)~dxdy\,,
\end{eqnarray*}
where $a$,$a^*$ are the usual creation-annihilation operator-valued distributions in the Fock space over $L^2(\mathbb{R}^d)$.
Recall that $a$ and $a^*$ satisfy the canonical commutation relations
$$
[a(x),a^*(y)]=\delta(x-y)\,,\;\; [a^*(x),a^*(y)]=0=[a(x),a(y)]\,.
$$
A simple computation leads to the following identity
$$
\varepsilon^{-1}H_{\varepsilon_{|L^2_s(\mathbb{R}^{d\N})}}=\mathbf{\rm H}_\N , \;\;\mbox{ if } \varepsilon=\frac{1}{\N}\,.
$$
Thus, the statement on the chaos propagation stated in (\ref{correl}) may be written (up to an unessential factor) as
\begin{eqnarray*}
\lim_{\hbarr\to 0}\big\langle
e^{-it\varepsilon^{-1} H_\varepsilon} \Psi_\varepsilon^0\,,\,b^{Wick} e^{-it\varepsilon^{-1} H_\varepsilon} \Psi_\varepsilon^0\big\rangle
=\la \varphi_t^{\otimes k},\O\varphi_{t}^{\otimes k}\ra\,,
\end{eqnarray*}
where $b^{Wick}$ denotes $\varepsilon$-dependent Wick observables defined by
\bean
b^{Wick}=\varepsilon^{k} \int_{\mathbb{R}^{2kd}}\; \prod_{i=1}^k a^*(x_i) \;\;\tilde \O(x_1,\cdots,x_k;y_1,\cdots,y_k)\; \prod_{j=1}^k a(y_j)
\; dx_1\cdots dx_k dy_1\cdots dy_k,
\eean
with $\tilde \O(x_1,\cdots,x_k;y_1,\cdots,y_k)$  the distribution kernel of a bounded operator $\O$ on $L^{2}(\mathbb{R}^{kd})$.
Therefore, the mean field limit $\N\to \infty$ for $\mathbf{\rm H}_\N $ can be converted to a semiclassical limit
$\varepsilon \to 0$ for $H_\varepsilon$. The study of the semiclassical  limit of the many-boson systems traces back to
the work of Hepp \cite{Hep} and was subsequently  improved by Ginibre and Velo \cite{GiVe1,GiVe2}. The latter analysis are based on
coherent states, {\it i.e.},
$$
\Psi_\varepsilon^0=e^{-\frac{|\varphi|^{2}}{2\hbarr}} \sum_{n=0}^\infty  \hbarr^{-n/2}
\frac{\varphi^{\otimes n}}{\sqrt{n!}}\,,\;\; \varphi\in L^{2}(\mathbb{R}^d)\,,
$$
which have infinite number of particles in contrast to the Hermite states $\Psi_\N^0=\varphi_0^{\otimes \N}$.
However,  a simple argument in the work of Rodnianski and Schlein \cite{RodSch} shows that  the semiclassical
analysis is enough to justify the chaos conservation hypothesis and even provides convergence estimates on
the $k$-particle correlation functions. The authors of \cite{RodSch} considered the problem under the assumption
of $(-\Delta+1)^{1/2}$-bounded potential ({\it i.e.}, $ V(-\Delta+1)^{-1/2}$ is bounded). The main purpose of the
present paper is to extend the latter result to more singular potentials using the ideas
of Ginibre and Velo \cite{GiVe2}.

\bigskip

For the sake of clarity, we restrict ourselves in this paper  to the particular example of point interaction potential
in one dimension, {\it i.e.},
\bea
\label{pmdelta}
V(x)=\delta(x)\,,\quad x\in\rz\,.
\eea
This example is  typical for potentials  which are
$-\Delta$-form bounded  ({\it i.e.}, $(-\Delta+1)^{-1/2}V(-\Delta+1)^{-1/2}$ is bounded). Indeed, we believe that such
simple example sums up the principal difficulties on the problem. Moreover,  we state  in Appendix \ref{appendix_sch}
some abstract results on the non-autonomous Schr\"odinger equation which have their
own interest and allow to consider a more general setting. We also remark that the results here can be easily
extended to the case $V(x)=-\delta(x)$ at the price to work locally in time.

\bigskip

The paper is organized as follows. We first recall the basic definitions for the Fock space framework in Section \ref{prem}. Then we
accurately introduce the quantum dynamics of the considered many-boson system and its classical counterpart, namely the cubic NLS equation.
The study of the semiclassical limit through Hepp's method is  carried out in Section \ref{Hepp} where we use results
on the time-dependent quadratic approximation derived in Section \ref{tquadsec}. Finally, in Section \ref{sec.chaos}
we apply the argument of \cite{RodSch} to prove the chaos propagation result.

\section{Preliminaries}
\label{prem}
Let $\mathfrak{H}$ be a Hilbert space. We denote by $\L(\mathfrak{H})$  the space of all linear bounded operators
on $\mathfrak{H}$.  For a linear unbounded operator $L$ acting on $\mathfrak{H}$, we denote by $\D(L)$ (
respectively $\mathcal{Q}(L)$) the operator domain (respectively form domain) of $L$. Let $D_{x_j}$ denotes the
differential operator $-i\partial_{x_j}$ on $L^2(\rz^n)$ where $(x_1,\cdots,x_n)\in\rz^n$.

\bigskip

In the following we recall the second quantization framework. We denote
by  $L_s^{2}(\mathbb{R}^{nd})$ the space of symmetric square integrable
functions, {\it i.e.},
\bean
\Psi_n\in L_s^{2}(\mathbb{R}^{nd})\;\;  \mbox{ iff } \; \; \Psi_n\in L^{2}(\mathbb{R}^{nd}) \; \;\mbox{ and }\; \;
\Psi_n(x_1,\cdots,x_n)=\Psi_n(x_{\sigma_1},\dots,x_{\sigma_n}) \quad \mbox{ a.e.},
\eean
for any permutation $\sigma$ on the symmetric group ${\rm Sym}(n)$.
The orthogonal projection from $ L^2(\rz^{nd})$ onto the closed subspace $ L_s^2(\rz^{nd})$ is given by
$$
\mathfrak{S}_{n}\Psi_n(x_1,\cdots,x_n)=\frac{1}{n!}\sum_{\sigma\in {\rm Sym}(n)}\Psi_n(x_{\sigma(1)},\cdots,x_{\sigma(n)}),\;\;\;
\Psi_n \in L^2(\rz^{nd})\,.
$$
We will often use the notation
$$
\S_s(\rz^{nd}):=\mathfrak{S}_{n} \S(\rz^{nd})\,
$$
where $\S(\rz^{nd})$ is the Schwartz space on $\rz^{nd}$. The symmetric Fock space over $L^2(\mathcal{\rz})$ is defined
as the Hilbert space,
\begin{eqnarray*}
\mathcal{F}=\bigoplus_{n=0}^\infty L_s^2(\rz^{nd})\,,
\end{eqnarray*}
endowed with the inner product
\bean
\la \Psi,\Phi\ra=\sum_{n=0}^\infty \int_{\rz^{nd}} \overline{\Psi_n(x_1,\cdots,x_n)}  \; \Phi_n(x_1,\cdots,x_n) \;
dx_1\cdots dx_n\,,
\eean
where  $\Psi=(\Psi_n)_{n\in\nz}$ and $\Phi=(\Phi_n)_{n\in\nz}$ are two arbitrary vectors in $\F$. A convenient subspace of $\F$  is
given as the algebraic direct sum
$$
\S:=\bigoplus_{n=0}^{\rm alg}\S_s(\rz^{nd})\,.
$$
Most essential linear operators on $\F$ are determined by their action  on the  family of vectors
\bean
\varphi^{\otimes n}(x_1,\dots,x_n)= \prod_{i=1}^n \varphi(x_i)\,, \;\;\; \varphi\in L^2(\rz^d)\,, 
\eean
which spans the  space $ L_s^2(\rz^{nd})$ thanks to the polarization identity,
\bean
\mathfrak{S}_n \prod_{i=1}^n \varphi_i(x_i)=\ds \frac{1}{2^n n!}
\sum_{\varepsilon_i=\pm 1} \varepsilon_1\cdots \varepsilon_n \;
\prod_{i=1}^n \big( \sum_{j=1}^n \varepsilon_j
\varphi_j(x_i)\big)\,.
\eean
For example, the creation and annihilation operators  $a^*(f)$ and $a(f)$, parameterized by $\hbarr>0$,
are defined by
\begin{eqnarray*}
a(f) \varphi^{\otimes n}&=&\sqrt{\hbarr n} \; \;\la f,\varphi\ra \varphi^{\otimes (n-1)}\\
a^*(f)\varphi^{\otimes n}&=&\sqrt{\hbarr (n+1)}  \;\;\mathfrak{S}_{n+1}
(\;f\otimes \varphi^{\otimes n})\,,\;\;\, \forall \varphi, f\in L^2(\rz^d).
\end{eqnarray*}
They can also by written as
\bean
a(f)=\sqrt{\hbarr}\int_{\rz^d} \overline{f(x)} \, a(x)\,dx, \quad a^*(f)=\sqrt{\hbarr}\int_{\rz^d} f(x) \, a^*(x)\,dx, 
\eean
where $a^*(x),a(x)$ are the canonical creation-annihilation operator-valued distributions.
Recall that for any $\Psi=(\Psi^{(n)})_{n\in\nz}\in\S$, we have
\bean
&&[a(x)\Psi]^{(n)}(x_1,\cdots,x_n)=\sqrt{(n+1)} \Psi^{(n+1)}(x,x_1,\cdots,x_n),\\
&& [a^*(x)\Psi]^{(n)}(x_1,\cdots,x_n)=\frac{1}{\sqrt{n}} \sum_{j=1}^n \delta(x-x_j) \Psi^{(n-1)}(x_1,\cdots,\hat x_j,\cdots, x_n)\,,
\eean
where $\delta$ is the Dirac distribution at the origin and  $\hat x_j$ means that the variable $x_j$ is omitted.
The Weyl operators are given for $f\in L^2(\rz^d)$  by
$$
W(f)=e^{\frac{i}{\sqrt{2}} [a^*(f)+a(f)]}\,,
$$
and they satisfy the Weyl commutation relations,
\begin{eqnarray}
\label{eq.Weylcomm}
W(f_1) W(f_2)=e^{-\frac{i\hbarr}{2} {\rm Im}\langle f_1, f_2\rangle} \;W(f_1+f_2),
\end{eqnarray}
with $f_1,f_2\in L^2(\rz^d)$.

\bigskip

Let us briefly recall the Wick-quantization procedure of polynomial symbols.
\begin{defn}
We say that a function $b:\S(\rz^d)\to \cz$  is a continuous $(p,q)$-homogenous polynomial
on $\S(\rz^d)$ iff it satisfies:\\
(i) $b(\lambda z)=\bar\lambda^q\lambda^p b(z)$ for any $\lambda\in\mathbb{C}$ and $z\in \S(\rz^d)$, \\
(ii) there exists a (unique) continuous hermitian form $\mathfrak{Q}:\S_s(\rz^{dq})\times \S_s(\rz^{dp}) \to \cz$ such that
$$
b(z)=\mathfrak{Q}(z^{\otimes q},z^{\otimes p}).
$$
We denote by $\E$ the vector space spanned by all those polynomials.
\end{defn}
The Schwartz kernel theorem ensures for any continuous $(p,q)$-homogenous polynomial $b$,
the existence of a kernel $\tilde b(.,.)\in \S'(\rz^{d(p+q)})$ such that
\bean
b(z)= \int_{\rz^{d(p+q)}} \tilde b(k'_1,\cdots,k_q';k_1,\cdots,k_p) \, \overline{z(k_1')\cdots z(k_q')}\, z(k_1)\cdots z(k_p)\; dk'dk\,,
\eean
in the distribution sense. The set of $(p,q)$-homogenous polynomials $b\in\E$ such that the kernel
$\tilde b$ defines a  bounded  operator from $L_s^2(\rz^{d p})$ into $L_s^2(\rz^{d q})$ will be denoted by
$\P_{p,q}(L^2(\rz^d))$. Those classes of polynomial symbols are  studied and used in \cite{AmNi1,AmNi2}.
\begin{defn}
The Wick quantization is the map which associate to each continuous $(p,q)$-homogenous polynomial
$b\in\E$, a quadratic form $b^{Wick}$ on $\S$ given by
\bean
\ds
\la\Psi, b^{Wick} \Phi\ra &=& \hbarr^{\frac{p+q}{2}}\; \int_{\rz^{d (p+q)}} \;\tilde b(k',k)\; \hspace{.1in} \la a(k_1')\cdots a(k_q') \Psi,
a(k_1)\cdots a(k_p) \Phi\ra_\F \;\;dk\;dk' \\
&=& \sum_{n=p}^\infty \hbarr^{\frac{p+q}{2}} \frac{\sqrt{n! (n-p+q)!}}{(n-p)!} \,
\int_{\rz^{d(n-p)}}dx\,\int_{\rz^{d(p+q)}}dkdk'\, \tilde b(k',k) \, \overline{\Psi^{(n)}(k,x)} \Phi^{(n-p+q)}(k',x),
\eean
for any $\Phi,\Psi\in\S$.
\end{defn}
We have, for example,
\begin{eqnarray*}
a^*(f)=\la z,f\ra^{Wick}   \hspace{.1in} \mbox{ and } \hspace{.1in}   a(f)=\la f, z\ra^{Wick} \,.
\end{eqnarray*}
Furthermore, for any self-adjoint operator $A$ on $L^2(\rz^d)$ such that $\S(\rz^d)$ is a core for $A$, 
the Wick quantization
$$
\d\Gamma(A):=\la z,A z\ra^{Wick}\,,
$$
defines a self-adjoint operator on $\F$. In particular, if $A$ is the identity we get the $\hbarr$-dependent
number operator
$$
N:=\la z, z\ra^{Wick}\,.
$$
We recall the standard number estimate (see, {\it e.g.}, \cite[Lemma 2.5]{AmNi1}),
\bea
\label{number_est}
\left|\la \Psi,b^{Wick}\Phi\ra\right|\leq ||\tilde b||_{\L(L_s^2(\rz^{dp}),L_s^2(\rz^{dq}))}\;||N^{q/2}\Psi|| \times ||N^{p/2}\Phi||\,,
\eea
which holds uniformly in $\hbarr\in(0,1]$ for $b\in\P_{p,q}(L^2(\rz^d))$ and any $\Psi,\Phi\in\D(N^{max(p,q)/2})$.

\section{Many-boson system}
\label{boson}
In nonrelativistic many-body theory, boson systems are  described by the second quantized Hamiltonian in
the symmetric Fock space  $\F$  formally given by
\begin{eqnarray}
\label{hamiltonian}
-\hbarr \int_{\rz^d} a^*(x) \Delta a(x) dx+\frac{\hbarr^2}{2} \int_{\rz^d}\int_{\rz^d} a^*(x)a^*(y) \delta(x-y) a(x)a(y) \,dxdy\,.
\end{eqnarray}
The rigorous meaning of formula (\ref{hamiltonian}) is as a quadratic form on $\S$, which we denote by $h^{Wick}$, obtained by
Wick quantization of  the classical energy functional
\bea
\label{energy-func}
   h(z)= \int_{\rz^d} |\nabla z(x)|^2 \, dx+P(z), \quad \mbox{ where } \quad
   P(z)= \frac{1}{2} \int_{\rz^d} |z(x)|^4 \, dx,\;\;\;z\in \S(\rz^d)\,.
\eea
More explicitly, we have for $\Psi\in\S$
\bean
\la\Psi, h^{Wick} \Psi\ra &=& \hbarr \sum_{n=1}^\infty n \int_{\rz^{dn}} \left|\partial_{x_1}\Psi^{(n)}(x_1,\cdots,x_n)\right|^2
dx_1\cdots dx_n \\
& +&\hbarr^2\sum_{n=2}^\infty \frac{n(n-1)}{2} \int_{\rz^{d(n-1)} } \left|
\Psi^{(n)}(x_2,x_2,\cdots,x_n)\right|^2 \, dx_2\cdots dx_n\,.
\eean
Moreover, in one dimensional space ({\it i.e.}, $d=1$) one can show the existence of a unique self-adjoint operator
bounded from below, which we denote by $H_\hbarr$, such that 
$$
\la\Psi,H_\hbarr\Psi\ra=\la\Psi,h^{Wick}\Psi\ra,\quad \mbox{ for  any } \; \Psi\in\S.
$$
This is proved in  Proposition \ref{selfadj}.

In all the sequel we restrict our analysis to space \underline{dimension $d=1$} and consider the small parameter
$\hbarr$ such that \underline{ $\hbarr\in(0,1]$}. The $\hbarr$-independent self-adjoint operator,
$$
S_\mu\Psi:= \Psi+\sum_{n=1}^\infty \left[n^\mu \,\Psi^{(n)}+\sum_{j=1}^n -\Delta_{x_j}\Psi^{(n)} \right]=\left(\hbarr^{-1}\d\Gamma(-\Delta)+
\hbarr^{-\mu} N^\mu +1\right)\Psi\,,
$$
with $\mu>0$, defines the Hilbert space $\F_+^{\mu}$ given as the linear space $\D(S_\mu^{1/2})$ equipped with the
inner product
$$
\la\Psi,\Phi\ra_{\F_{+}^{\mu}}:=\la S_\mu^{1/2}\Psi, S_\mu^{1/2}\Phi\ra_\F.
$$
We denote by $\F_{-}^{\mu}$ the completion  of $\D(S_\mu^{-1/2})$ with respect to the norm associated to the following  inner product
$$
\la\Psi,\Phi\ra_{\F_-^\mu}:=\la S^{-1/2}_\mu\Psi, S^{-1/2}_\mu\Phi\ra_\F.
$$
Therefore, we have the Hilbert rigging
$$
\F_+^\mu\subset \F\subset \F_-^\mu.
$$
Note that the form domain of the $\varepsilon$-dependent self-adjoint  operator $\d\Gamma(-\Delta)+N^\mu$ with $\mu>0$ is
$$
\mathcal{Q}(\d\Gamma(-\Delta)+N^\mu)=\F_+^\mu \quad \mbox{ for any } \varepsilon\in(0,1]\,.
$$

\begin{lem} For any $\Psi,\Phi\in\S$,
\label{est.Q}
\bean
\left|\la \Psi, P^{Wick} \Phi\ra \right| \leq \frac{1}{4} \,||[\d\Gamma(-\Delta)+N^3]^{1/2}\Psi||
\;\times\;||[\d\Gamma(-\Delta)+N^3]^{1/2} \Phi||\,.
\eean
\end{lem}
\proof
A simple computation yields for any $\Psi,\Phi\in\S$
\bean
\la\Psi, P^{Wick}\Phi\ra= \sum_{n=2}^\infty \varepsilon^2 \frac{n (n-1)}{2} \int_{\rz^{n-1}}
\overline{\Psi^{(n)}(x_2,x_2,x_3,\cdots,x_n)} \;\Phi^{(n)}(x_2,x_2,x_3,\cdots,x_n)\,
dx_2\cdots dx_n\,.
\eean
Cauchy-Schwarz inequality yields
\bean
\left|\la\Psi, P^{Wick}\Phi\ra\right| &\leq& \left[\sum_{n=2}^\infty \varepsilon^2 \frac{n (n-1)}{2} \int_{\rz^{n-1}}
|\Psi^{(n)}(x_2,x_2,x_3,\cdots,x_n)|^2 \,dx_2\cdots dx_n \right]^{1/2} \\
&\times& \; \left[\sum_{n=2}^\infty \varepsilon^2 \frac{n (n-1)}{2} \int_{\rz^{n-1}}
|\Phi^{(n)}(x_2,x_2,x_3,\cdots,x_n)|^{2}\,dx_2\cdots dx_n \right]^{1/2}\,.
\eean
Using Lemma \ref{main-est}, we get for any $\alpha(n)>0$
\bean
\left|\la\Psi, P^{Wick}\Phi\ra\right| &\leq& \left[\sum_{n=2}^\infty\varepsilon^2 \frac{n (n-1)}{2\sqrt{2}} \left(
\alpha(n) \la D_{x_1}^2\Psi^{(n)},\Psi^{(n)}\ra+\frac{\alpha(n)^{-1}}{2}\la \Psi^{(n)},\Psi^{(n)}\ra\right)\right]^{1/2}\\
&\times&\, \left[\sum_{n=2}^\infty\varepsilon^2 \frac{n (n-1)}{2\sqrt{2}} \left(
\alpha(n) \la D_{x_1}^2\Phi^{(n)},\Phi^{(n)}\ra+\frac{\alpha(n)^{-1}}{2}\la \Phi^{(n)},\Phi^{(n)}\ra\right)\right]^{1/2}.
\eean
Hence, by choosing $\alpha(n)=\frac{1}{\sqrt{2}\varepsilon (n-1)}$, it follows that
\bean
\left|\la\Psi, P^{Wick}\Phi\ra\right| &\leq &\frac{1}{4} \left[\sum_{n=2}^\infty \varepsilon n \la D_{x_1}^2\Psi^{(n)},\Psi^{(n)}\ra+
\sum_{n=2}^\infty \varepsilon^3 n(n-1)^2 \la\Psi^{(n)},\Psi^{(n)}\ra\right]^{1/2}\\
&&\times \left[\sum_{n=2}^\infty \varepsilon n \la D_{x_1}^2\Phi^{(n)},\Phi^{(n)}\ra+
\sum_{n=2}^\infty \varepsilon^3 n(n-1)^2 \la\Phi^{(n)},\Phi^{(n)}\ra\right]^{1/2}\\
&\leq& \frac{1}{4}  \sqrt{\la\Psi,[\d\Gamma(-\Delta)+N^3] \Psi\ra} \, \times \, \sqrt{\la \Phi,[\d\Gamma(-\Delta)+N^3] \Phi\ra}\;.
\eean
This leads to the claimed estimate.\hfill\cqfd

\begin{remark}
Note that, as in Lemma \ref{est.Q}, the estimate
\bea
\label{est.Q-epsi}
\left|\la \Psi, P^{Wick} \Phi\ra \right| \leq \frac{\hbarr^2}{4} \,||\Psi||_{\F_+^3} \;\;||\Phi||_{\F_+^3}\,
\eea
holds true for any $\Psi,\Phi\in\S$ and $\hbarr\in(0,1]$.
\end{remark}

We can show that $h^{Wick}$ is associated to a self-adjoint operator by considering its restriction to each
sector $L_s^2(\rz^n)$, however we will prefer the following point of view.

\begin{prop}
\label{selfadj}
There exists a unique  self-adjoint operator $H_{\hbarr}$ such that
$$
\la\Psi, h^{Wick} \Phi\ra= \la \Psi, H_{\hbarr} \Phi\ra \; \mbox{ for any } \; \Psi\in\F_+^3, \Phi\in\D(H_\hbarr)\cap\F_+^3\,.
$$
Moreover, $e^{-it/\hbarr H_\hbarr}$ preserves $\F_+^3$.
\end{prop}
\proof
We first use the KLMN theorem (\cite[Theorem X17]{RS}) and Lemma \ref{est.Q}  to show that the quadratic form
$h^{Wick}+N^3+1$ is associated to a unique (positive) self-adjoint operator $L$ with
$$
\mathcal{Q}(L)=\mathcal{Q}(\d\Gamma(-\Delta)+N^3)=\F_+^3\,.
$$
Observe that we also have
\bea
\label{equi-norms}
||[\d\Gamma(-\Delta)+N^3]^{1/2}\Psi||\leq ||L^{1/2} \Psi|| \mbox{ for any } \Psi\in \F_+^3\,.
\eea
Next, by the Nelson commutator theorem (Theorem \ref{nelson}) we can prove that the quadratic form $h^{Wick}$ is
uniquely associated to a self-adjoint operator denoted by $H_\varepsilon$ with
$\D(L)\subset \D(H_\hbarr)\cap\F_+^3$  and deduce the invariance of $\F_+^3$. Indeed,
we easily check using Lemma \ref{est.Q} and (\ref{equi-norms}) that
\bea
\label{n1}
\left|\la \Psi, h^{Wick}\Phi\ra \right|\leq \frac{5}{4} \; ||L^{1/2}\Psi|| \; \;||L^{1/2}\Phi||\, \mbox{ for any } \Psi,\Phi\in\F_+^3.
\eea
Furthermore, we have for $\Psi,\Phi\in\F_+^3$ and $\lambda>0$
\bea
\label{n2}
\la L(\lambda L+1)^{-1}\Psi, h^{Wick}(\lambda L+1)^{-1}\Phi\ra -\la (\lambda L+1)^{-1}\Psi, h^{Wick} L (\lambda L+1)^{-1}\Phi\ra =0\,.
\eea
The statements (\ref{n1})-(\ref{n2}) with the help of Lemma \ref{equi-assmp-nels}, allow to use Theorem \ref{nelson}. \hfill\cqfd
\begin{remark}
\label{opeg}
The same argument as in Proposition \ref{selfadj} shows that the quadratic form on $\F_+^3$ given by
\bean
G:=\hbarr^{-1} \d\Gamma(-\Delta)+\hbarr^{-2}P^{Wick}+\hbarr^{-1}N+1\,,
\eean
is associated to a unique (positive) self-adjoint operator which we denote by the same symbol $G$.
\end{remark}

\section{The cubic NLS equation}
Let $H^s(\rz^m)$ denote the Sobolev spaces. The energy functional $h$ given by
(\ref{energy-func}) has the associated vector field
\bean
X:H^1(\rz)&\longrightarrow& H^{-1}(\rz)\\
z&\longmapsto & X(z)=-\Delta z+\partial_{\bar z} P(z)\,,
\eean
which leads to the nonlinear classical field equation
\bea
\label{nls}
\begin{array}{lcl}
i \partial_t \varphi&=& X(\varphi)\\
&=&-\Delta \varphi+ |\varphi|^2 \varphi
\end{array}
\eea
with initial data $\varphi_{|t=0}=\varphi_0 \in H^{1}(\rz)$. It is well-known that the
above cubic defocusing NLS equation  is globally well-posed on $H^{s}(\rz)$ for $s\geq 0$.
In particular,  the equation (\ref{nls}) admits a unique global solution on $C^0(\rz,H^{m}(\rz))\cap
C^1(\rz,H^{m-2}(\rz))$  for any initial data $\varphi\in H^{m}(\rz)$ when  $m=1$  and $m=2$ (see \cite{GiVe3} for
$m=1$ and \cite{T} for $m=2$). Moreover, we have energy and mass  conservations {\it i.e.},
\bean
h(\varphi_t)=h(\varphi_0) \quad \mbox{ and } \quad ||\varphi_t||_{L^2(\rz)}=||\varphi_0||_{L^2(\rz)}\,,
\eean
for any initial data $\varphi_0\in H^1(\rz)$ and $\varphi_t$ solution of (\ref{nls}).
It is not difficult to prove the following estimates
\bea
\label{energy-est}
\begin{array}{ccccc}
||\varphi||_{L^{\infty}(\rz)}^2&\leq & 2 ||\varphi||_{L^{2}(\rz)} \; ||\partial_x\varphi||_{L^{2}(\rz)}  & \leq
& \,2  \,||\varphi||_{L^{2}(\rz)}\;h(\varphi)^{1/2}\,,\\ \nm
||\varphi||_{L^p(\rz)}^p &\leq&  2^{\frac{p-2}{2}} ||\varphi||_{L^{2}(\rz)}^{\frac{p+2}{2}} \;
||\partial_x\varphi||_{L^{2}(\rz)}^{\frac{p-2}{2}}
&\leq& 2^{\frac{p-2}{2}} \;||\varphi||_{L^{2}(\rz)}^{\frac{p+2}{2}}\,\,h(\varphi)^{\frac{p-2}{4}}\,,
\end{array}
\eea
for $p\geq 2$ and any $\varphi\in H^1(\rz)$. Furthermore, using Gronwall's inequality we show for any $\varphi_0\in H^2(\rz)$
the existence of  $c>0$ depending only on $\varphi_0$  such that
\bea
\label{H2-est}
||\varphi_t||_{H^2(\rz)}\leq e^{c\,|t|\,} \;\;||\varphi_0||_{H^2(\rz)}\,,
\eea
where  $\varphi_t$ is a solution  of the NLS equation (\ref{nls}) with initial condition $\varphi_0$.

\section{Time-dependent quadratic dynamics}
\label{tquadsec}
In this section we construct a time-dependent quadratic approximation for the Schr\"odinger  dynamics.
We prove existence of a unique unitary propagator for this approximation using the abstract results for non-autonomous linear
Schr\"odinger  equation stated in the Appendix \ref{appendix_sch}. This step will be useful for the study of propagation of coherent states
in the semiclassical limit in section \ref{sec-coherent}.

\bigskip
\noindent
The polynomial $P$ has the following Taylor expansion for any $z_0\in H^1(\rz)$
\bean
P(z+z_0)=\sum_{j=0}^4 \frac{D^{(j)}P}{j!}(z_0)[z]\,.
\eean
Let $\varphi_t$ be a solution of the NLS equation (\ref{nls}) with an initial data $\varphi_0\in H^{1}(\mathbb{R})$.
Consider the time-dependent quadratic polynomial on $\S(\mathbb{R})$ given by
\bean
P_2(t)[z]&:=&\frac{D^{(2)}P}{2}(\varphi_t)[z]\\
&=& {\rm Re}\int_{\rz} \overline{z(x)}^2 \, \varphi_t(x)^2\, dx +2 \int_{\mathbb{R}} |z(x)|^2 \,|\varphi_t(x)|^2\,dx\,.
\eean
Let  $\{A_2(t)\}_{t\in \rz}$ be the $\hbarr$-independent family of quadratic forms on $\S$ defined by
\bea
\label{quad-ham}
\hbarr A_2(t):=\d\Gamma(-\Delta)+P_2(t)^{Wick}\,.
\eea

\begin{lem}
\label{scalequad}
For $\varphi_0\in H^1(\mathbb{R})$ let
$$
\vartheta_1:=16^2 (||\varphi_0||_{L^2(\rz)}+1)^3(h(\varphi_0)+1) \;\mbox{ and  }\;
\vartheta_2:=16^2 (||\varphi_0||_{L^2(\rz)}+1)^{3/2}\sqrt{h(\varphi_0)+1}.
$$
The  quadratic forms on $\S$ defined by
$$
S_2(t):=A_2(t)+ \vartheta_1 \hbarr^{-1} N+\vartheta_2 1\,,\quad\;\;t\in \rz\,,
$$
are associated to unique self-adjoint operators, still denoted by $S_2(t)$, satisfying
\begin{itemize}
 \item $S_2(t)\geq  1$,
 \item $\D(S_2(t)^{1/2})=\F_+^1$ for any $\;t\in\rz$\,.
\end{itemize}
\end{lem}
\proof
The case $\varphi_0=0$ is trivial.
By definition of Wick quantization we have for $\Psi,\Phi\in\S$,
\bea
\label{comp-p2}
\begin{array}{l}
\ds\la \Phi,P_2(t)^{Wick} \Psi\ra=2\sum_{n=1}^\infty  \hbarr n \int_{\mathbb{R}^n } |\varphi_t(x_1)|^2 \,
\overline{\Phi^{(n)}(x_1,\cdots,x_n)} \Psi^{(n)}(x_1,\cdots,x_n)\,dx_1\cdots dx_n\\ \ds
+ \ds\sum_{n=0}^\infty  \hbarr \sqrt{(n+1)(n+2)} \int_{\mathbb{R}^n} \overline{\Phi^{(n)}(x_1,\cdots,x_n)} \,\left(\int_{\mathbb{R}}
\overline{\varphi_t(x)}^2 \Psi^{(n+2)}(x,x,x_1,\cdots,x_n) dx\right)dx_1\cdots dx_n\,\\
+ \ds\sum_{n=0}^\infty  \hbarr \sqrt{(n+1)(n+2)} \int_{\mathbb{R}^n} \Psi^{(n)}(x_1,\cdots,x_n) \,\left(\int_{\mathbb{R}}
\varphi_t(x)^2 \overline{\Phi^{(n+2)}(x,x,x_1,\cdots,x_n)} dx\right)dx_1\cdots dx_n.
\end{array}
\eea
Therefore, using Cauchy-Schwarz inequality, we show
\bean
|\la \Phi,P_2(t)^{Wick} \Psi\ra|&\leq&  2||\varphi_t||_{L^{\infty}(\mathbb{R})}^2 || N^{1/2}\Phi||\times ||N^{1/2}\Psi|| \\
&+& ||\varphi_t||_{L^4(\rz)}^2 \, ||(N+\hbarr)^{1/2}\Phi||\times  \left[\sum_{n=0}^\infty \hbarr (n+2)
||\Psi^{(n+2)}(x,x,x_1,\cdots,x_n)||_{L^2(\rz^{n+1})}^2\right]^{1/2} \\
&+& ||\varphi_t||_{L^4(\rz)}^2 \, ||(N+\hbarr)^{1/2}\Psi||\times  \left[ \sum_{n=0}^\infty \hbarr (n+2)
||\Phi^{(n+2)}(x,x,x_1,\cdots,x_n)||_{L^2(\rz^{n+1})}^2\right]^{1/2}\,.
\eean
Now we prove, by Lemma \ref{main-est}, the crude estimate
\bean
|\la \Phi,P_2(t)^{Wick} \Psi\ra| &\leq&
\max(||\varphi_t||^2_{L^4(\mathbb{R})},  ||\varphi_t||_{L^{\infty}(\mathbb{R})}^2)\;\ds
 \left[2 ||N^{1/2}\Phi|| \times ||N^{1/2}\Psi|| \right.\\
 &+&   ||(N+\hbarr)^{1/2} \Phi||\times
  ||(\alpha\d\Gamma(-\Delta)+\alpha^{-1} N)^{1/2}
 \Psi|| \\ &+& \left.||(N+\hbarr)^{1/2} \Psi||\times
  ||(\alpha\d\Gamma(-\Delta)+\alpha^{-1} N)^{1/2}
 \Phi||  \right]\,.
\eean
This yields for any $\alpha>0$
\bea
\label{est.pg}
\begin{array}{lll}
|\la \Phi,P_2(t)^{Wick} \Psi\ra| &\leq& \alpha \; \max(||\varphi_t||^2_{L^4(\mathbb{R})},
||\varphi_t||_{L^{\infty}(\mathbb{R})}^2)\;\ds
\\ &&\times ||\left[\d\Gamma(-\Delta)+(\alpha^{-1}+3)\alpha^{-1} N
+ \alpha^{-1} \hbarr 1\right]^{1/2}\Phi|| \\
&&\times ||\left[\d\Gamma(-\Delta)+(\alpha^{-1}+3)\alpha^{-1} N
+ \alpha^{-1}\hbarr 1\right]^{1/2}\Psi||\,.
\end{array}
\eea
Remark now that (\ref{energy-est}) yields
$$
\max(||\varphi_t||^2_{L^4(\mathbb{R})},||\varphi_t||_{L^{\infty}(\mathbb{R})}^2)\leq 2 \,(||\varphi_0||_{L^2(\rz)}+1)^{3/2}
\sqrt{h(\varphi_0)+1}\,.
$$
Hence, for  $\alpha^{-1}=3 (||\varphi_0||_{L^2(\rz)}+1)^{3/2}\sqrt{h(\varphi_0)+1}>0$, we obtain
\bea
\label{est.p2t}
\begin{array}{ccc}
\hbarr^{-1}|\la \Phi,P_2(t)^{Wick} \Psi\ra|&\leq &
\frac{2}{3}  \; ||[\hbarr^{-1}\d\Gamma(-\Delta)+\vartheta_1\hbarr^{-1}N+\vartheta_21]^{1/2}\Phi|| \\ & &
\times ||[\hbarr^{-1}\d\Gamma(-\Delta)+\vartheta_1\hbarr^{-1}N+\vartheta_2 1]^{1/2}\Psi||\,.
\end{array}
\eea
Applying now the KLMN theorem (see \cite[Theorem X.17]{RS}) with the help of inequality (\ref{est.p2t}) we
show that
$$
S_2(t)=A_2(t)+\vartheta_1 \hbarr^{-1} N+ \vartheta_2 1 \, \mbox{ with } \vartheta_1>(\alpha^{-1}+3)\alpha^{-1}, \mbox{ and }
\vartheta_2>\alpha^{-1}+1\,,
$$
are associated to  unique self-adjoint operators $S_2(t)$ satisfying $S_2(t)\geq 1$. Furthermore, we have that the form domains
of those operators are time-independent, {\it i.e.},
$$
\mathcal{Q}(S_2(t))=\F_+^1
$$
for any $t\in\rz$.\hfill\cqfd

\begin{remark}
The choice of $\vartheta_1$, $\vartheta_2$ in the previous lemma takes into account the use of KLMN's theorem in 
the proof of Lemma \ref{coh_lem3}.
\end{remark}

\bigskip
We consider the non-autonomous  Schr\"odinger equation
\bea
\label{bos-quad}
\left\{
 \begin{array}[c]{l}
   i\partial_t u=A_2(t) u\,,\quad t\in \mathbb{R},\\
   u(t=s)=u_s\,.
 \end{array}
\right.
\eea
Here $\rz\ni t\mapsto A_2(t)$ is considered as a norm continuous  $\L(\F_+^1,\F_-^1)$-valued  map (see Lemma \ref{asump1}).
We show in Proposition \ref{quad-propag} the existence of a unique solution for any initial data $u_s\in\F_+^1$ using Corollary \ref{abs_schrod_cor}.
Moreover, the Cauchy problem's features allow to encode the solutions on a {\it unitary propagator} mapping 
$(t,s)\mapsto U_2(t,s)$ such that
$$
U_2(t,s)u_s=u_t\,,
$$
satisfying Definition \ref{unit-propag} with $\H=\F$, $\H_\pm=\F_\pm^1$ and $I=\rz$.

\bigskip
\noindent
In the following two lemmas we check the assumptions in Corollary \ref{abs_schrod_cor}.
\begin{lem}
\label{asump1}
For any  $\varphi_0\in H^1(\mathbb{R})$ and $t\in\rz$ the quadratic form $A_2(t)$ defines a symmetric operator on
$\L(\F_+^1,\F_-^1)$ and the mapping $t\in \rz\mapsto A_2(t)\in\L(\F_+^1,\F_-^1)$ is norm continuous.
\end{lem}
\proof
Using (\ref{est.p2t}) we show for any $\Psi,\Phi\in\S$
\bea
\label{equi-norm2}
\begin{array}{lll}
|\la \Phi, A_2(t)\Psi\ra|&\leq & |\la \Phi, \hbarr^{-1}\d\Gamma(-\Delta)\Psi\ra|+|\la \Phi, \hbarr^{-1}P_2(t)^{Wick}\Psi\ra|\\ \nm\ds
&\leq& ||S_1^{1/2}\Phi|| \; ||S_1^{1/2}\Psi||+\frac{2}{3} \vartheta_1 ||S_1^{1/2}\Phi|| \; ||S_1^{1/2}\Psi|| \\ \nm\ds
&\leq& \frac{5}{3} \, \vartheta_1 \;||\Psi||_{\F_+^1} \;||\Phi||_{\F_+^1}\,,
\end{array}
\eea
where $\vartheta_1,\vartheta_2$ are the parameters introduced in Lemma \ref{scalequad}. Hence, this allows to consider $A_2(t)$ as a bounded operator
in $\L(\F_+^1,\F_-^1)$. Since $A_2(t)$ is a symmetric quadratic form it follows that it is also symmetric  as an operator in
$\L(\F_+^1,\F_-^1)$. \\
Now, using a similar estimate as (\ref{est.pg}) we prove norm continuity. Indeed, we have
\bean
|\la \Phi, [A_2(t)-A_2(s)]\Psi\ra|&=& \hbarr^{-1}|\la \Phi, [P_2(t)-P_2(s)]^{Wick}\Psi \ra|\\
&\leq& 4\;
\max\left(||\varphi_t^2-\varphi_s^2||_{L^2(\mathbb{R})},\left\||\varphi_t|^2-|\varphi_s|^2\right\|_{L^{\infty}(\mathbb{R})}\right)
\; ||\Psi||_{\F_+^1} \;||\Phi||_{\F_+^1}\,.
\eean
Note that it is not difficult to prove that
$$
\max\left(||\varphi_t^2-\varphi_s^2||_{L^2(\mathbb{R})},\left\||\varphi_t|^2-|\varphi_s|^2\right\|_{L^{\infty}(\mathbb{R})}\right)
\ds\longrightarrow 0\,\quad \mbox{ when } t\to s\,.
$$
This follows by (\ref{energy-est}) and the fact that $\varphi_t\in C^0(\rz,H^1(\rz))$.
\hfill\cqfd

\begin{lem}
\label{asump2}
For any $\varphi_0\in H^2(\mathbb{R})$ there exists $c>0$ (depending only on $\varphi_0$) such that the two statements below hold true.\\
(i) For any $\Psi\in \F_+^1$, we have
\bean
|\partial_t\la \Psi,  S_2(t) \Psi\ra |\leq e^{c (|t|+1)} ||S_2(t)^{1/2}\Psi||_{\F}\,.
\eean
(ii) For any $\Psi,\Phi\in\D(S_2(t)^{3/2})$, we have
\bean
\left| \la\Psi, A_2(t) S_2(t) \Phi\ra-\la S_2(t)\Psi,A_2(t) \Phi\ra \right| \leq c \, ||S_2(t)^{1/2}\Psi||_{\F} \;
||S_2(t)^{1/2}\Phi||_{\F}\,.
\eean
\end{lem}
\proof
(i) Let $\Psi\in\S$, we have
\bean
\partial_t \la \Psi,S_2(t) \Psi\ra&=&\hbarr^{-1}\; \partial_t \la \Psi, P_2(t)^{Wick} \Psi\ra\\
&=&  \hbarr^{-1}\la \Psi, \left[\partial_t P_2(t)\right]^{Wick} \Psi\ra\,,
\eean
where $\partial_t P_2(t)$ is a continuous polynomial on $\S(\rz)$ given by
\bean
\partial_t P_2(t)[z]=2 {\rm Re}\int_{\rz} \overline{z(x)}^2 \, \varphi_t(x)\partial_t\varphi_t(x)\, dx +4
{\rm Re}\int_{\mathbb{R}} |z(x)|^2 \,\overline{\varphi_t(x)}\partial_t\varphi_t(x)\,dx\,.
\eean
A simple computation yields
\bean
&&\la \Psi, \left[\partial_t P_2(t)\right]^{Wick} \Psi\ra = 4{\rm Re}\; \sum_{n=1}^\infty  n \hbarr \;
\overbrace{\int_{\rz^{n}}\, \ds
\overline{\varphi_t(x_1)} \partial_t\varphi_t(x_1) \, |\Psi^{(n)}(x_1,\cdots,x_n)|^2 \,dx_1\cdots dx_n}^{(1)}\\
&&+\sum_{n=0}^\infty \hbarr\sqrt{(n+2)(n+1)} \int_{\rz^n} \overline{\Psi^{(n)}(x_1,\cdots,x_n)} \left(
\int_\rz \overline{\varphi_t(x)} \,\overline{\partial_t\varphi_t(x)}\,
\Psi^{(n+2)}(x,x,x_1,\cdots,x_n)\;dx\right) dx_1\cdots dx_n\\
&&+hc\,.
\eean
From (\ref{energy-est}) we get
\bean
\left|(1)\right| &\leq& ||\varphi_t \,\partial_t\varphi_t||_{L^1(\rz)} \; \ds\int_{\rz^{n-1}} \;\sup_{x_1\in\rz}
\left|\Psi^{(n)}(x_1,\cdots,x_n)\right|^2 dx_2\cdots dx_n\\
&\leq&  ||\varphi_t||_{L^2(\rz)}\times ||\partial_t\varphi_t||_{L^2(\rz)}\;
\la (1-\partial_{x_{1}}^2)\Psi^{(n)},\Psi^{(n)}\ra_{L^2(\rz^n)}\,.
\eean
Now we apply Cauchy-Schwarz inequality,
\bean
|\la \Psi,\left[\partial_t P_2(t)\right]^{Wick} \Psi\ra|&\leq & 4\;||\varphi_t||_{L^2(\rz)}\; ||\partial_t\varphi_t||_{L^2(\rz)}
\left(\sum_{n=1}^\infty \hbarr n \,\la (1-\partial_{x_{1}}^2)\Psi^{(n)},\Psi^{(n)}\ra_{L^2(\rz^n)}\right)
\\&& \hspace{-1.5in}+ 2 ||\varphi_t||_{L^\infty(\rz)} ||\partial_t\varphi_t||_{L^2(\rz)}\left(\sum_{n=0}^\infty \hbarr \ds (n+2) ||\Psi^{(n+2)}(x,x,.)||_{L^2(\rz^{n+1})}^2\right)^{1/2} \times
\left(\sum_{n=0}^\infty \hbarr \ds (n+1) ||\Psi^{(n)}||^2_{L^2(\rz^n)}\right)^{1/2}.
\eean
In the same spirit as in (\ref{est.pg}), we obtain a rough inequality
\bean
|\la \Psi,\left[\partial_t P_2(t)\right]^{Wick} \Psi\ra|&\leq&
\max(||\varphi_t||_{L^\infty(\rz)},||\varphi_t||_{L^2(\rz)})\; ||\partial_t\varphi_t||_{L^2(\rz)} \; \left[4\,||(\d\Gamma(-\Delta)+N)^{1/2} \Psi||^2 \right.\\
&&+ 2\; \left.||(\d\Gamma(-\Delta)+N+1)^{1/2} \Psi||^2\right]\,.
\eean
Observe that (\ref{est.p2t}) implies  $S_1\leq 3\,S_2(t)$ for all $t\in\rz$.
Hence, we have
\bean
\hbarr^{-1}|\la \Psi,\left[\partial_t P_2(t)\right]^{Wick} \Psi\ra| &\leq& 6\, \max(||\varphi_t||_{L^\infty(\rz)},||\varphi_t||_{L^2(\rz)})\; ||\partial_t\varphi_t||_{L^2(\rz)}
\; ||\Psi||_{\F_+^1}^2\\
&\leq & 18 \,\max(||\varphi_t||_{L^\infty(\rz)},||\varphi_t||_{L^2(\rz)})
\; ||\partial_t\varphi_t||_{L^2(\rz)}\;
||S_2(t)^{1/2}\Psi||_{\F}^2\,.
\eean
This proves (i) since (\ref{energy-est})-(\ref{H2-est}) ensure the existence of $c>0$ (depending only on $\varphi_0$) such that
\bean
\max(||\varphi_t||_{L^\infty(\rz)},||\varphi_t||_{L^2(\rz)})
\; ||\partial_t\varphi_t||_{L^2(\rz)}\leq e^{c(|t|+1)}\,.
\eean
(ii) If $\Psi,\Phi\in\D(S_2(t)^{3/2})$ the quantity
\bean
\mathcal{C}:=\la\Psi, A_2(t) S_2(t) \Phi\ra-\la S_2(t)\Psi,A_2(t) \Phi\ra,
\eean
is well-defined since $A_2(t)\in\L(\F_+^1,\F_-^1)$ and $S_2(t)\D(S_2(t)^{3/2})\subset\D(S_2(t)^{1/2})=\F_+^1$.
Note that $N\in\L(\F_+^1,\F_-^1)$. Hence, we can write
\bean
\mathcal{C}&=&\ds\la \Psi, [S_2(t)-\vartheta_1 \hbarr^{-1} N-\vartheta_2 1] \,S_2(t)  \Phi\ra
- \la S_2(t)  \Psi, [S_2(t)-\vartheta_1 \hbarr^{-1} N-\vartheta_2 1] \,\Phi\ra\\ \nm
&=& \ds \vartheta_1  \left(\la S_2(t) \Psi, \hbarr^{-1}N \Phi\ra-\la \hbarr^{-1}N \Psi, S_2(t) \,\Phi\ra\right)\,.
\eean
Observe that, for $\lambda>0$, $\hbarr^{-1}N (\lambda \hbarr^{-1}N+1)^{-1}\F_+^1\subset \F_+^1$ and that
$$
s-\lim_{\lambda\to 0^+} \hbarr^{-1}N
(\lambda \hbarr^{-1}N+1)^{-1}=\hbarr^{-1}N  \mbox{ in } \L(\F_+^1, \F_-^1).
$$
Therefore, we have
\bean
\mathcal{C}&=& \ds \vartheta_1  \lim_{\lambda\to 0^+} \underbrace{
\la S_2(t) \Psi, \hbarr^{-1}N(\lambda \hbarr^{-1}N+1)^{-1} \Phi\ra-\la \hbarr^{-1}N (\lambda\hbarr^{-1} N+1)^{-1} \Psi, S_2(t) \Phi\ra}_{\mathcal{C}_\lambda}\,.
\eean
Let $N_\lambda$ denote $\hbarr^{-1}N(\lambda \hbarr^{-1}N+1)^{-1}$. A simple computation yields
\bean
\hbarr \mathcal{C}_\lambda &=& \la  \Psi,   P_2(t)^{Wick} N_\lambda  \,\Phi\ra
- \la N_\lambda \, \Psi,  P_2(t)^{Wick} \Phi\ra\\
&=&\la   \Psi, g(t)^{Wick} N_\lambda \Phi\ra-\la N_\lambda \Psi, g(t)^{Wick} \Phi\ra \,,
\eean
where $g(t)$ is the polynomial given by
$$
g(t)[z]={\rm Re}\int_{\rz} \overline{z(x)}^2\; \varphi_t(x)^2 \,dx\,.
$$
A similar computation as (\ref{comp-p2}) yields
\bean
\begin{array}{lll}
\ds  \mathcal{C}_\lambda &=&  \ds\sum_{n=0}^\infty  \kappa(n) \ds\;
\int_{\mathbb{R}^n} \overline{\Psi^{(n)}(x_1,\cdots,x_n)} \,\left(\int_{\mathbb{R}}
\overline{\varphi_t(x)}^2 \Phi^{(n+2)}(x,x,x_1,\cdots,x_n) dx\right)dx_1\cdots dx_n\,
\\ \nm
&&- \ds\sum_{n=0}^\infty  \kappa(n) \ds\;
\int_{\mathbb{R}^n} \Phi^{(n)}(x_1,\cdots,x_n) \,\left(\int_{\mathbb{R}}
\varphi_t(x)^2 \overline{\Psi^{(n+2)}(x,x,x_1,\cdots,x_n)} dx\right)dx_1\cdots dx_n\,,
\end{array}
\eean
where
$$
\kappa(n)=\frac{(n+2)\sqrt{(n+1)(n+2)}}{(\lambda (n+2)+1)}- \frac{n\sqrt{(n+1)(n+2)}}{(\lambda n+1)}\,.
$$
Note that $\kappa(n)\leq 2 (n+2)$. Hence, using Cauchy-Schwarz inequality, we show
\bean
\left| \mathcal{C}_\lambda \right|&\leq& 2\, ||\varphi_t||^2_{L^4(\rz)}\;
\left[\sum_{n=0}^\infty (n+2) ||\Psi^{(n)}||_{L^2(\rz^n)}^2\right]^{1/2}
\left[\sum_{n=0}^\infty (n+2) \,||\Phi^{(n+2)}(x,x,.)||_{L^2(\rz^{n+1})}^2\right]^{1/2}
\\ \nm \ds&&+2\,||\varphi_t||^2_{L^4(\rz)}\;
\left[\sum_{n=0}^\infty (n+2) ||\Phi^{(n)}||_{L^2(\rz^n)}^2\right]^{1/2}
\left[\sum_{n=0}^\infty (n+2) \,||\Psi^{(n+2)}(x,x,.)||_{L^2(\rz^{n+1})}^2\right]^{1/2}\,.
\eean
Using Lemma \ref{main-est}, with $\alpha=\frac{1}{\sqrt{2}}$, we get
\bean
\sum_{n=0}^\infty (n+2) \,||\Psi^{(n+2)}(x,x,.)||_{L^2(\rz^{n+1})}^2
&\leq& \frac{1}{2}\;\sum_{n=0}^\infty (n+2) \la D_{x_1}^2 \Psi^{(n+2)},\Psi^{(n+2)}\ra + (n+2) ||\Psi^{(n+2)}||_{L^2(\rz^{n+2})}^2\\
&\leq&\frac{1}{2}\;  \la \Psi, S_1 \Psi\ra\,,
\eean
together with an analogue  estimate where $\Psi$ is replaced by $\Phi$.
Now, we conclude that there exists $c>0$ depending only on $\varphi_0$ such that
\bea
\label{clambda}
\vartheta_1 \left|\mathcal{C}_\lambda \right|&\leq&  c \;||\Psi||_{\F_+^1} \,||\Phi||_{\F_+^1}\,.
\eea
This proves part (ii).\hfill\cqfd

\begin{prop}
\label{quad-propag}
Let $\varphi_0\in H^2(\rz)$ and $A_2(t)$ given by (\ref{quad-ham}). Then the non-autonomous Cauchy problem
\bean
\left\{
 \begin{array}[c]{l}
   i\partial_t u=A_2(t) u\,,\quad t\in \mathbb{R},\\
   u(t=s)=u_s\,,
 \end{array}
\right.
\eean
admits a unique unitary propagator $U_2(t,s)$ in the sense of Definition \ref{unit-propag}
with $I=\rz$ and $\H_\pm=\F_\pm^1$. Moreover, there exists $c>0$ depending only on $\varphi_0$ such that
\bean
||U_2(t,0)||_{\L(\F_+^1)}\leq e^{c e^{c|t|}}\,.
\eean
\end{prop}
\proof
The proof immediately follows using Corollary \ref{abs_schrod_cor} with the help of
Lemma \ref{asump1}-\ref{asump2} and the inequality
$$
c_1 S_1\leq S_2(t)\leq c_2 S_1,
$$
which holds true using (\ref{equi-norm2}).\hfill\cqfd

\section{Propagation of coherent states}
\label{Hepp}
In finite dimensional phase-space, coherent state analysis is a well developed powerful tool, see for instance
\cite{CRR}. Here we study, using the ideas of Ginibre and Velo in \cite{GiVe2},
the asymptotics when $\hbarr\to 0$ of the time-evolved coherent states
\bean
e^{-it/\hbarr H_\hbarr} W(\frac{\sqrt{2}}{i\hbarr} \varphi_0)\Psi\,,
\eean
for $\Psi$ in a dense subspace $\G_+\subset \F$ defined below. We consider the following Hilbert rigging
$$
\G_+\subset \F\subset \G_-\,,
$$
defined via the $\hbarr$-independent self-adjoint operator (see Remark \ref{opeg}) given by
$$
G:=\hbarr^{-1}\d\Gamma(-\Delta)+\hbarr^{-2} P^{Wick}+\hbarr^{-1} N +1\,,
$$
as the completion of $\D(G^{\pm 1/2})$ with the respect to the inner product
$$
\la\Psi,\Phi\ra_{\G_\pm}:=\la G^{\pm 1/2}\Psi, G^{\pm 1/2}\Phi\ra_\F.
$$
We have the continuous embedding
\bean
\F_+^3\subset \G_+\subset \F_+^1\,.
\eean
The main result of this section is the following proposition which describes the propagation of coherent states in the
semiclassical limit.
\label{sec-coherent}
\begin{prop}
\label{coherent}
For any $\varphi_0\in H^2(\rz)$ there exists $c>0$  depending  only on $\varphi_0$  such that
$$
\left\|e^{-i t/\hbarr H_{\hbarr}} W(\frac{\sqrt{2}}{i\hbarr}\varphi_0)\Psi-e^{i\omega(t)/\hbarr}
W(\frac{\sqrt{2}}{i\hbarr}\varphi_{t})U_{2}(t,0)\Psi\right\|_{\F}\leq
e^{c e^{c|t|}}\;\hbarr^{1/8}\;\left\|\Psi\right\|_{\G_+}\,,
$$
holds for any $t\in\rz$ and $\Psi\in\G_+$ where $\varphi_t$  solves the NLS equation (\ref{nls}) with the initial condition $\varphi_0$  and
$\omega(t)=\int_{0}^{t} P(\varphi_{s})~ds$. Here $U_2(t,s)$ is the unitary propagator given by Proposition \ref{quad-propag}.
\end{prop}

\bigskip

To prove this proposition we need several preliminary lemmas.
\begin{lem}
\label{diff-weyl}
The  following three assertions hold true.
\begin{description}
\item (i) For any $\xi\in L^2(\rz)$ and $k\in\nz$, the Weyl operator $W(\xi)$ preserves $\D(N^{k/2})$. If in addition
$\xi\in H^1(\rz)$ then $W(\xi)$ preserves also $\F_+^\mu$ when $\mu\geq 1$.
\item (ii) For any  $\xi\in H^1(\rz)$, we have in the sense of quadratic forms on  $\F_+^3\,,$
\bean
W(\frac{\sqrt{2}}{i\hbarr} \xi)^* \; h^{Wick}\; W(\frac{\sqrt{2}}{i\hbarr} \xi)=h(.+\xi)^{Wick}\,.
\eean
\item (iii)  Let $(\rz\ni t\mapsto \varphi_t)\in C^1(\rz,L^2(\rz))$, then for any $\Psi\in\D(N^{1/2})$ we have in $\F$
 \bean
i \hbarr \partial_t W(\frac{\sqrt{2}}{i\hbarr} \varphi_t)\Psi&=& W(\frac{\sqrt{2}}{i\hbarr} \varphi_t) \left[
{\rm Re}\la \varphi_t, i\partial_t \varphi_t\ra+2{\rm Re}\la z,i\partial_t\varphi_t\ra^{Wick}\right]\Psi\\
&=& \left[- {\rm Re}\la \varphi_t, i\partial_t \varphi_t\ra+2{\rm Re}\la z,i\partial_t\varphi_t\ra^{Wick}\right]W(\frac{\sqrt{2}}{i\hbarr} \varphi_t)\Psi\,.
\eean
\end{description}
\end{lem}
\proof
(i) Let $\F_0$ be the linear space spanned by vectors $\Psi\in\F$ such that
$\Psi^{(n)}=0$ for any $n$ except for a finite number. It is known that for any $\xi\in L^2(\rz)$ and $\Psi\in\F_0$
\bea
\label{wick-weyl}
\tilde N\Psi:=W(\frac{\sqrt{2}}{i\hbarr}\xi)^*N \,W(\frac{\sqrt{2}}{i\hbarr}\xi)\Psi=
\left(N+2 {\rm Re}\la z,\xi\ra^{Wick}+||\xi||^2 1\right)\Psi\,.
\eea
For a proof of the latter identity see \cite[Lemma 2.10 (iii)]{AmNi1}. Hence, by Cauchy-Schwarz inequality it follows that
\bean
||N^{1/2} W(\frac{\sqrt{2}}{i\hbarr}\xi)\Psi||^2 &=& \la \Psi,\left[N+ 2 {\rm Re}\la z,\xi\ra^{Wick}+||\xi||^2 1\right]\Psi\ra \\
&=& \la \Psi, (N+||\xi||_{L^2(\rz)}^2 1)\Psi\ra \\ &+&\sum_{n=0}^\infty\sqrt{\hbarr (n+1)} \;
\int_{\rz^n} \overline{\Psi^{(n)}(y)} \left(\int_\rz \overline{\xi(x)}  \Psi^{(n+1)}(x,y) dx \right)dy+hc
\\ &\leq& (1+||\xi||_{L^2(\rz)})^2 \; ||(N+1)^{1/2} \Psi||^2\,.
\eean
Now, for $k\geq 1$ we show the existence of an $\hbarr$-independent constant $C_k>0$ depending only on
$k$ and $||\xi||_{L^2(\rz)}$ such that
\bea
\label{numberequiv}
||N^{k/2} W(\frac{\sqrt{2}}{i\hbarr}\xi)\Psi||^2=\la \Psi,\tilde N^k\Psi\ra  \leq C_k \;||(N+1)^{k/2} \Psi||^2\,.
\eea
This is a consequence of the number operator estimate (\ref{number_est}) and the fact that $\tilde N^k$ is a Wick polynomial in $\sum_{0\leq r,s\leq k}\P_{r,s}(L^2(\rz))$ (see, {\it e.g.},\cite[Prop.~2.7 (i)]{AmNi1}). Thus, we have proved
the invariance of $\D(N^{k/2})$ since $\F_0$ is a core of $N^{k/2}$.

Now the invariance of $\F_+^\mu$, $\mu\geq 1$,  follows by Faris-Lavine Theorem \ref{farislavine} where we take the operator
$$
A=\sqrt{2}{\rm Re}\la z,\xi\ra^{Wick} \hspace{.2in}\mbox{ and }\hspace{.2in}  S=S_\mu=\hbarr^{-1}\d\Gamma(-\Delta)+\hbarr^{-\mu} N^\mu+1\,,
$$
and remember that
\bean
W(\xi)=e^{i\sqrt{2}{\rm Re}\la z,\xi\ra^{Wick}}\,.
\eean
In fact, assuming $\xi\in H^1(\rz)$ we have to check
assumptions (i)-(ii) of  Theorem \ref{farislavine}.
For any $\Psi\in \F_+^\mu$, we have by Wick quantization
\bean
2{\rm Re}\la z,\xi\ra^{Wick}\Psi&=&\sum_{n=0}^\infty \sqrt{\hbarr (n+1)} \,
\int_{\rz} \overline{\xi(x)} \Psi^{(n+1)}(x,x_1,\cdots,x_n) \,dx \\&&+ \sum_{n=1}^\infty
\sqrt{\frac{\hbarr}{n}} \sum_{j=1}^n \xi(x_j) \, \Psi^{(n-1)}(x_1,\cdots,\hat x_j,\cdots,x_n)\,.
\eean
Therefore, it is easy to show
\bean
||{\rm Re}\la z,\xi\ra^{Wick}\Psi||&\leq &  \sqrt{\hbarr}||\xi||_{L^2(\rz)}\, ||(\hbarr^{-1}N+1)^{1/2}\Psi||\\
&\leq &  \sqrt{\hbarr}||\xi||_{L^2(\rz)}\, ||S_1\Psi||\,,
\eean
and hence we obtain that  $\D(S_\mu)\subset\D(A)$. Let $\Psi\in \D(S_\mu)$, a standard computation yields
\bea
\label{comtat}
\begin{array}{lll}
\sqrt{2}\left(\la A \Psi, S_\mu\Psi\ra-\la S_\mu\Psi, A\Psi\ra\right)&=&\la a(-\Delta \xi) \Psi,\Psi\ra- \la \Psi, a(-\Delta \xi)\Psi\ra
\\ \nm\ds &+&\la [(\frac{N}{\hbarr}+1)^\mu-(\frac{N}{\hbarr})^\mu]\, \Psi, a^*(\xi)\Psi\ra- hc\,.
\end{array}
\eea
Each two terms in the same line of (\ref{comtat}) are similar and it is enough to estimate only one of them. We have
by Cauchy-Schwarz inequality
\bean
\left|\la a(-\Delta \xi) \Psi,\Psi\ra \right|&\leq & \left|
\sum_{n=0}^\infty \sqrt{\hbarr (n+1)} \int_{\rz^n} \overline{
\Psi^{(n)}(y)}\left(
\int_\rz \overline{-\Delta \xi(x)} \Psi^{(n+1)} (x,y) dx \right) dy\right| \\ &\leq & ||\xi||_{H^1(\rz)} \; ||S_1^{1/2} \Psi||^2\,,
\eean
and for $\,1\leq \theta\leq\mu-1$
\bean
\left|\la \hbarr^{-\theta}N^\theta \Psi, a^*(\xi)\Psi\ra\right|&\leq & \left|\sum_{n=0}^\infty \sqrt{ \hbarr (n+1)} \, (n+1)^\theta
\int_{\rz^n} \Psi^{(n)}(y)\left(
\int_\rz \xi(x) \overline{\Psi^{(n+1)} (x,y)} dx \right) dy\right| \\
&\leq & 2^\mu||\xi||_{L^2(\rz)}\,||S^{1/2}_\mu\Psi||^2\,.
\eean
This shows for any $\Psi\in\D(S_\mu)$,
\bean
\pm i \la\Psi, [A,S_\mu]\Psi\ra\leq C \; ||S_\mu^{1/2}\Psi||^2.
\eean
Part (ii) follows by a similar argument as \cite[Lemma 2.10 (iii)]{AmNi1} and part (iii) is a well-known formula, see \cite[Lemma 3.1 (3)]{GiVe1}.\hfill\cqfd

\bigskip
\noindent
Set
\bean
\W(t)=W(\frac{\sqrt{2}}{i\hbarr} \varphi_t)^* \, e^{-i\omega(t)/\hbarr} \, e^{-it/\hbarr H_\hbarr}
W(\frac{\sqrt{2}}{i\hbarr} \varphi_0)\,.
\eean

\begin{lem}
\label{coh_lem3}
For any $\varphi_0\in H^2(\rz)$ there exists $c>0$ such that the inequality
\bean
\left\| \W(t) \right\|_{\L(\G_+,\F_+^1)}\leq e^{c e^{c|t|}} \,
\eean
holds  for $t\in\rz$ uniformly in $\hbarr\in(0,1]$.
\end{lem}
\proof
Observe that the subspace $\D_+$ given as the image of $\D(H_\hbarr)\cap\F_+^3$ by $
W(\frac{\sqrt{2}}{i\hbarr} \varphi_0)^*$ is dense in $\F$. Let $\Psi\in\D_+$
and $\Phi\in\G_+$, then differentiating the quantity $\la\Phi,\W(t)\Psi\ra$  with the help of Lemma \ref{diff-weyl} and
Proposition \ref{selfadj}, we obtain
\bea
\label{eqtime4}
\begin{array}{lll}
i\hbarr \partial_t \la\Phi,\W(t)\Psi\ra&=& \la\Phi,\, 
[P(\varphi_t)-{\rm Re}\la \varphi_t, i\partial_t \varphi_t\ra-2 {\rm Re}\la z,i\partial_t\varphi_t\ra^{Wick}]\,
\W(t)\Psi\ra\\ \nm\ds
&+& \ds\underbrace{\la \Phi, W(\frac{\sqrt{2}}{i\hbarr} \varphi_t)^* \, e^{-i\omega(t)/\hbarr} H_\hbarr \,W(\frac{\sqrt{2}}{i\hbarr} \varphi_0)\Psi\ra}_{
(1)}\,.
\end{array}
\eea
Let $R_\nu:=1_{[0,\nu]}(\hbarr^{-1}N)$ and remark that $s-\lim_{\nu\to\infty}R_\nu=1$. Furthermore, we have that $R_\nu\G_+\subset\F_+^3$ since
it easily holds that
\bean
||R_\nu\Phi||_{\F_+^3}^2\leq \nu^3 \,||\Phi||_{\G_+}^2\,.
\eean
Therefore, since $W(\frac{\sqrt{2}}{i\hbarr} \varphi_t) R_\nu\Phi$ and $W(\frac{\sqrt{2}}{i\hbarr} \varphi_0)\Psi$ belong to $\F_+^3$, we have
\bean
(1)&=&\lim_{\nu\to\infty} \la R_\nu\Phi, W(\frac{\sqrt{2}}{i\hbarr} \varphi_t)^* \, e^{-i\omega(t)/\hbarr} H_\hbarr
\,W(\frac{\sqrt{2}}{i\hbarr} \varphi_0)\Psi\ra\\
&=&\lim_{\nu\to\infty} \la R_\nu\Phi, h(.+\varphi_t)^{Wick} \,\W(t)\Psi\ra\,.
\eean
So, we get
\bean
\begin{array}{lll}
i\hbarr \partial_t \la\Phi,\W(t)\Psi\ra&=& (1)+
\ds\lim_{\nu\to\infty}
\ds\la R_\nu\Phi,\left[P(\varphi_t)-{\rm Re}\la \varphi_t, i\partial_t \varphi_t\ra-2 {\rm Re}\la z,i\partial_t\varphi_t\ra^{Wick}\right]
\W(t) \Psi\ra \\ \nm\ds
&=&\ds\lim_{\nu\to\infty}
\la R_\nu\Phi, (\underbrace{\hbarr A_2(t)+P_3(t)^{Wick}+P^{Wick}}_{=:\hbarr\Theta(t)})\W(t)\,\Psi\ra\,,
\end{array}
\eean
where we denote
\bean
P_3(t)[z]:=\frac{D^{(3)}P}{3!} (\varphi_t)[z]=2{\rm Re} \int_\rz \varphi_t(x) \overline{z(x)} |z(x)|^2 \,dx\,\hspace{.1in} \mbox{ and } \hspace{.1in}
P(z)=\frac{D^{(4)}P}{4!}(\varphi_t)[z]=\frac{1}{2}\int_\rz |z(x)|^4 \, dx\,.
\eean
A simple computation yields
\bean
\la \Phi,P_3(t)^{Wick}\Psi\ra &= &\sum_{n=1}^\infty \sqrt{n^2(n+1)\hbarr^{3}}
\int_{\rz^{n-1}} \left(\int_{\rz} \overline{\varphi_t(x)} \,\overline{\Phi^{(n)}(x,y)} \,\Psi^{(n+1)}(x,x,y) \,dx\right) dy
\\ \nm \ds &&+\sum_{n=1}^\infty \sqrt{n^2(n+1)\hbarr^{3}}
\int_{\rz^{n-1}} \left(\int_{\rz} \varphi_t(x) \,\overline{\Phi^{(n+1)}(x,x,y)} \,\Psi^{(n)}(x,y) \,dx\right) dy\,.
\eean
Using Cauchy-Schwarz inequality and Lemma \ref{main-est}, we obtain
\bea
\label{p3est}
\begin{array}{lll}
\left|\la \Phi,P_3(t)^{Wick}\Psi\ra \right|&\leq& 2 \sqrt{2} \frac{||\varphi_t||_{L^{\infty}(\rz)}}{\sqrt{\vartheta_2}} \;
\sqrt{\la\Phi,[\hbarr^{-1}P^{Wick}+\vartheta_1 \hbarr^{-1}N+\vartheta_2 1]\Phi\ra} \\ \nm\ds
&\times& \sqrt{\la\Psi,[\hbarr^{-1}P^{Wick}+\vartheta_1 \hbarr^{-1}N+\vartheta_2 1]\Psi\ra}\,,
\end{array}
\eea
where $\vartheta_1,\vartheta_2$ are the parameters in Lemma \ref{scalequad}.
Hence, $\Theta(t)$ extends to a bounded operator in $\L(\G_+,\G_-)$ since  $A_2(t)$ and $P^{Wick}$ belong to
$\L(\G_+,\G_-)$. As an immediate consequence we obtain
\bea
\label{derw}
i\hbarr \partial_t \la\Phi,\W(t)\Psi\ra=\la\Phi,\hbarr \Theta(t) \W(t)\Psi\ra.
\eea
Now, we consider the quadratic form $\Lambda(t)$  on $\G_+$ given by
\bean
\Lambda(t):=\Theta(t)+\vartheta_1 \hbarr^{-1} N+\vartheta_2 1\,.
\eean
It is easily follows, by (\ref{energy-est}) and (\ref{p3est}), that
\bea
\label{2oneinter}
\begin{array}{lll}
\left|\la \Phi,P_3(t)^{Wick}\Psi\ra \right|&\leq& \frac{1}{4} \;
||\left(-\hbarr^{-1}\d\Gamma(-\Delta)+\hbarr^{-1}P^{Wick}+\vartheta_1\hbarr^{-1}N+\vartheta_2 1\right)^{1/2}\Phi||\\
& &  \;\;||\left(-\hbarr^{-1}\d\Gamma(-\Delta)+\hbarr^{-1}P^{Wick}+\vartheta_1\hbarr^{-1}N+\vartheta_2 1\right)^{1/2}\Psi||\,.
\end{array}
\eea
Therefore, using  (\ref{est.p2t}) and (\ref{2oneinter}) we show that
\bean
\hbarr^{-1}\left[\frac{D^{(2)}P}{2}(\varphi_t)[z]+\frac{D^{(3)}P}{3!} (\varphi_t)[z]\right]^{Wick}
\eean
is  form bounded by $\hbarr^{-1}\d\Gamma(-\Delta)+\hbarr^{-1}P^{Wick}+\vartheta_1\hbarr^{-1}N+\vartheta_21$ with a form-bound less than $1$
uniformly in $\hbarr\in(0,1]$.
Hence, by the KLMN Theorem \cite[Thm.~X17]{RS},  the quadratic form $\Lambda(t)$
is associated to a unique self-adjoint operator which we still denote by $\Lambda(t)$, satisfying $\mathcal{Q}(\Lambda(t))=\G_+$ and $ \Lambda(t)\geq 1$.
Moreover, it is not difficult to show the existence of $c_1,c_2>0$ such that
\bea
\label{lem3-eq.1}
c_1 \, S_1\leq \Lambda(t) \leq  c_2 \,G
\eea
uniformly in $\hbarr\in(0,1]$ for any $t\in\rz$ .
Now, we consider the non-autonomous Schr\"odinger equation
\bea
\label{wschrod}
i\partial_t u_t =\Theta(t) u_t\,,
\eea
with initial data $u_0\in\G_+$. Next, we prove existence and uniqueness of a unitary propagator $\V(t,s)$ of the Cauchy 
problem (\ref{wschrod}). This will be done if we can check assumptions
of Corollary \ref{abs_schrod_cor} with $\G_\pm=\H_\pm$, $A(t)=\Theta(t)$
and $S(t)=\Lambda(t)$. Thus, we will conclude that
\bea
\label{eq.estq3}
||\Lambda(t)^{1/2}\V(t,0)\Psi||_\F\leq e^{c e^{c|t|}} \; ||\Lambda(0)^{1/2}\Psi||_\F\,.
\eea
Observe that $\rz\ni t\mapsto \Theta(t)\in\L(\G_+,\G_-)$ is norm continuous since
\bean
|\la\Phi,(\Theta(t)-\Theta(s))\, \Psi\ra|\leq ||\Phi||_{\G_+} \,||A_2(t)-A_2(s)||_{\L(\F_+^1,\F_-^1)}\,||\Psi||_{\G_+}+|
\la \Phi,\hbarr^{-1}(P_3(t)-P_3(s))^{Wick} \;\Psi\ra|\,,
\eean
and an estimate  similar to (\ref{p3est}) yields
\bean
|\la \Phi,\hbarr^{-1}(P_3(t)-P_3(s))^{Wick} \;\Psi\ra|\leq 2\sqrt{2} ||\varphi_t-\varphi_s||_{L^\infty(\rz)} \, ||\Phi||_{\G_+}\, ||\Psi||_{\G_+}\,.
\eean
Let us check assumption (i) of Corollary \ref{abs_schrod_cor}. We have for $\Psi\in\G_+\subset\F_+^1$,
\bean
\partial_t\la \Psi, \Lambda(t) \Psi\ra=\partial_t\la \Psi, S_2(t)\Psi\ra+\partial_t\la \Psi,
\hbarr^{-1} P_3(t)^{Wick} \Psi\ra\,.
\eean
A simple computation yields
\bean
\partial_t\la\Psi,\hbarr^{-1} P_3(t)^{Wick} \Psi\ra=2 {\rm Re}\left[
\sum_{n=1}^\infty \sqrt{n^2 (n+1) \hbarr} \int_{\rz^{n-1}}\left(\int_\rz \partial_t\varphi_t(x)
\Psi^{(n)}(x,y) \overline{\Psi^{(n+1)}(x,x,y)} dx\right)dy\right]\,.
\eean
So, by Cauchy-Schwarz inequality and Lemma \ref{main-est}, we get
\bean
\left|\partial_t \la \Psi,\hbarr^{-1} P_3(t)^{Wick} \Psi\ra\right|&\leq&
2 ||\partial_t\varphi_t||_{L^2(\rz)} \, \left[\sum_{n=1}^\infty (n+1) ||\sup_{x\in\rz} \left|\Psi^{(n)}(x,.)\right|||_{L^{2}(\rz^{n-1})}^2\right]^{1/2} \\
&\times& \left[\sum_{n=1}^\infty n^2\hbarr ||\Psi^{(n+1)}(x,x,.)||^2_{L^2(\rz^n)}\right]^{1/2}\\
&\leq& 2\sqrt{2} \,||\partial_t\varphi_t||_{L^2(\rz)} \, || \Lambda(t)^{1/2} \Psi||^2\,.
\eean
The latter estimate with Lemma \ref{asump2} (i) and (\ref{energy-est})-(\ref{H2-est}) give us
\bean
\left|\partial_t \la \Psi, \Lambda(t) \Psi\ra\right|\leq e^{c (|t|+1)} ||\Lambda(t)^{1/2} \Psi||^2\,.
\eean
Now, we check assumption (ii) of Corollary \ref{abs_schrod_cor}. We follow the same lines of the proof of Lemma \ref{asump2} (ii) by replacing $S_2(t)$ by $\Lambda(t)$ and
$A_2(t)$ by $\Theta(t)$. So, we arrive at the step where we have to estimate for $\Psi,\Phi\in\D(\Lambda(t)^{3/2})$ and $\lambda>0$, the quantity
\bean
\mathcal{C}_\lambda [g(t)] &:=&\la   \Psi, \hbarr^{-1}
g(t)^{Wick} N_\lambda \Phi\ra-\la N_\lambda \Psi, \hbarr^{-1} g(t)^{Wick} \Phi\ra \,,
\eean
where $N_\lambda:=\hbarr^{-1}N(\lambda \hbarr^{-1}N+1)^{-1}$ and $g(t)$ is the continuous  polynomial on $\S(\rz)$ given  by
\bean
g(t)[z]=P_2(t)[z]+P_3(t)[z]\,.
\eean
Note that the part $\mathcal{C}_\lambda[P_2(t)]$ involving only the symbol  $P_2(t)$ is already bounded by (\ref{clambda}).
Thus, we need only to consider $ \mathcal{C}_\lambda[P_3(t)]$.
A simple computation yields
\bean
\ds  \mathcal{C}_\lambda[P_3(t)] &=&  \ds\sum_{n=1}^\infty  \kappa(n) \ds\;
\int_{\mathbb{R}^{n-1}} \left(\int_\rz \overline{\varphi_t(x)}\, \Phi^{(n+1)}(x,x,y)\overline{\Psi^{(n)}(x,y)} \,
\,dx\right)\,dy\,
\\ \nm
&&- \ds\sum_{n=1}^\infty  \kappa(n) \ds\;
\int_{\mathbb{R}^{n-1}} \,\left(\int_{\mathbb{R}} \varphi_t(x)  \Phi^{(n)}(x,y)
\overline{\Psi^{(n+1)}(x,x,y)} \,dx\right) dy\,,
\eean
where
$$
\kappa(n)=\frac{(n+1)\sqrt{\hbarr n^2(n+1)}}{(\lambda (n+1)+1)}- \frac{n\sqrt{\hbarr n^2(n+1)}}{(\lambda n+1)}\,
$$
satisfying $|\kappa(n)|\leq  \sqrt{n^2(n+1)}$ uniformly in $\hbarr\in(0,1]$ and $\lambda>0$. So, using a similar estimate as
(\ref{p3est}), we obtain
\bean
\left| \mathcal{C}_\lambda[P_3(t)] \right|\leq \frac{1}{\sqrt{2}} \,||\varphi_t||_{L^\infty(\rz)}\; ||\Lambda(t)^{1/2}\Psi||\,
||\Lambda(t)^{1/2}\Phi||\,.
\eean
This proves assumption (ii) of Corollary \ref{abs_schrod_cor}. Now, we check that
$$
\W(t)=\V(t,0).
$$
In fact, for $\Phi\in\G_+$ and $\Psi\in\D_+$  we have
\bean
i\partial_r\la \Phi,\V(0,r)\W(r)\Psi\ra=-\la\Theta(r)\V(r,0)\Phi,\W(r)\Psi\ra+i\lim_{s\to 0} \la \V(r+s,0)\Phi,\frac{\W(r+s)-\W(r)}{s}
\Psi\ra \,,
\eean
and since by (\ref{eqtime4}) we know that $\lim_{s\to 0}\frac{\W(r+s)-\W(r)}{s}
\Psi$ exists in $\F$, we conclude using (\ref{derw}) that
\bean
\partial_r\la \Phi,\V(0,r)\W(r)\Psi\ra=0\,.
\eean
This identifies $\W(t)$ as the unitary propagator of the non-autonomous Schr\"odinger equation (\ref{wschrod}).
Therefore, by (\ref{lem3-eq.1})-(\ref{eq.estq3}) we get
\bean
\sqrt{c_1}\,||\W(t)\Psi||_{\F_+^1}\leq ||\Lambda(t)^{1/2}\W(t)\Psi||_{\F}\leq e^{c e^{c|t|}} ||\Lambda(0)^{1/2}\Psi||_{\F}\leq \sqrt{c_2}\, e^{c e^{c|t|}}||\Psi||_{\G_+}\,,
\eean
for any $t\in\rz$ uniformly in $\hbarr\in(0,1]$. \hfill\cqfd

\begin{lem}
\label{coh_lem2}
For any $\varphi_0\in H^2(\rz)$ and $\Psi\in\G_+$ we have
\bean
\left\|\W(t)\Psi-U_{2}(t,0)\Psi\right\|_{\F}^2&=&
2\la\Psi, (1-R_\nu)\Psi\ra -2{\rm Re}\la \W(t)\Psi,(1-R_\nu) U_2(t,0)\Psi\ra\\ \nm\ds&&
+2 \;\ds{\rm Im}\int_0^t \;\langle  \W(s)\Psi, \;[\Theta(s) R_\nu-R_\nu A_2(s)] \,U_{2}(s,0)\Psi\rangle \;ds\,,
\eean
where $R_\nu:=\sigma(\frac{\hbarr^{-1} N}{\nu})$ with $\sigma$ any bounded Borel function on $\rz_+$ with compact support and here
$$
\Theta(s)=A_2(s)+\hbarr^{-1} Q_s(z)^{wick}\,,
$$
with $Q_s(z)$ the continuous polynomial on $\S(\rz)$ given by
\bean
Q_s(z)&=&\frac{D^{(3)}P}{3!} (\varphi_s)[z]+\frac{D^{(4)}P}{4!}(\varphi_s)[z]\,.
\eean
\end{lem}
\proof
We have
\bea
\begin{array}{lll}
\label{lem2-eq.1}
\left\|\W(t)\Psi-U_{2}(t,0)\Psi\right\|_{\F}^2 &=&2\left\|\Psi\right\|^2_\F-
2 {\rm Re}\la \W(t)\Psi, U_{2}(t,0)\Psi\ra\,\\ \nm\ds &=&
2\la \Psi,(1-R_\nu)\Psi\ra-2 {\rm Re} \la \W(t)\Psi,(1-R_\nu) U_2(t,0)\Psi\ra \\
&&+2{\rm Re}\la \Psi,R_\nu \Psi\ra-2{\rm Re} \la \W(t)\Psi,R_\nu U_2(t,0)\Psi\ra.
\end{array}
\eea
Hence to prove the lemma it is enough to show that
\bea
\label{diff-coh}
\rz\ni s\mapsto {\rm Re} \la \W(s)\Psi,R_\nu U_2(s,0)\Psi\ra\in C^1(\rz)\,
\eea
and compute its  derivative.  Recall that the propagator $U_2(s,0)\in C^0(\rz,\L(\F_+^1))$,
by Proposition \ref{quad-propag} and that $\W(s)\in C^0(\rz,\L(\G_+))$ since it is the unitary propagator of
the Cauchy problem (\ref{wschrod}). It is easily seen that
\bean
s\mapsto R_\nu U_2(s,0)\Psi \,,
\eean
are in $\in C^0(\rz,\G_+)$ since $R_\nu$ maps continuously $\F_+^{1}$ into $\G_+$.  We also have that
\bean
s\mapsto \W(s)\Psi \in C^1(\rz,\G_-) \quad \mbox{ and }  \quad s\mapsto U_2(s,0)\Psi \in C^1(\rz,\F_-^1)\,.
\eean
This proves the statement (\ref{diff-coh}). Therefore, we have
\bea
\label{sec.coh.eq3}
2{\rm Re}\la \Psi,R_\nu \Psi\ra-2{\rm Re} \la \W(t)\Psi,R_\nu U_2(t,0)\Psi\ra=-\frac{2}{\hbarr} {\rm Im} \int_0^t
i\hbarr\partial_s\;\la \W(s)\Psi, R_\nu U_2(s,0)\Psi\ra \;ds\,.
\eea
The fact that $\W(t)$ is the unitary propagator of (\ref{wschrod})  with Proposition \ref{quad-propag} yields
\bea
\label{sec.coh.lem2eq1}
i\hbarr\partial_s\la \W(s)\Psi, R_\nu U_{2}(s,0)\Psi\ra =-\la \hbarr\Theta(s)\W(s)\Psi,R_\nu U_2(s,0)\Psi\ra
+\la \W(s)\Psi,  R_\nu \hbarr A_2(s)  U_{2}(s,0)\Psi\ra\,.
\eea
Now, collecting (\ref{lem2-eq.1}), (\ref{sec.coh.eq3}) and (\ref{sec.coh.lem2eq1}) we obtain the claimed identity.\hfill\cqfd

\bigskip
\noindent
{\bf Proof of Proposition~\ref{coherent}}
We are now ready to prove Proposition \ref{coherent}. \\
First observe that we have
\bean
\left\|e^{-i t/\hbarr H_{\hbarr}} W(\frac{\sqrt{2}}{i\hbarr}\varphi_0)\Psi-e^{i\omega(t)/\hbarr}
W(\frac{\sqrt{2}}{i\hbarr}\varphi_{t})U_{2}(t,0)\Psi\right\|_{\F}^2=
\left\|\W(t)\Psi-U_{2}(t,0)\Psi\right\|_{\F}^2\,.
\eean
Now, using Lemma  \ref{coh_lem2} one obtains for $t>0$ (the case $t<0$ is similar)
the estimate
\bean
\left\|\W(t)\Psi-U_{2}(t,0)\Psi\right\|_{\F}^2 &\leq & 2 \left|\la\Psi, (1-R_\nu)\Psi\ra\right|+ 2\left|
\la \W(t)\Psi,(1-R_\nu) U_2(t,0)\Psi\ra\right| \\ \nm\ds
&& +2 \;\ds\int_0^t \;\left|\langle  \W(s)\Psi, \;[\Theta(s) R_\nu-R_\nu A_2(s)] \,U_{2}(s,0)\Psi\rangle\right| \;ds\,.
\eean
Here we consider $\sigma$ to be in the class $C^1(\rz_+) $, decreasing and satisfying $\sigma(s)=1$ if $s\leq 1$ and   $\sigma(s)=0$ if $s\geq 2$.
We have for  $\nu$ positive  integer,
\bean
\label{5-est1}
\begin{array}{lllll}
\la\Psi,(1-R_\nu)\Psi\ra &\leq& \ds \frac{1}{\nu} \sum_{n=\nu+1}^\infty n \la  \Psi^{(n)},(D_{x_1}^2+1)\Psi^{(n)}\ra && \\ \nm\ds
&\leq& \frac{1}{\nu} \la \Psi, \hbarr^{-1}[\d\Gamma(-\Delta)+N]\Psi\ra&\leq &\frac{1}{\nu} \;\|\Psi\|_{\F_+^1}^2 \,.
\end{array}
\eean
Hence, we easily check with the help of Proposition \ref{quad-propag} and Lemma \ref{coh_lem3} that
\bean
\label{5-est2}
\begin{array}{lll}
\left|\la\W(t)\Psi,(1-R_\nu)U_2(t,0)\Psi\ra\right| &\leq& \frac{1}{\nu} \;||U_2(t,0)\Psi||_{\F_+^1}\;||\W(t)\Psi||_{\F_+^1} \\
&\leq& \frac{1}{\nu} \; e^{c_1 e^{c_1 t}} ||\Psi||_{\F_+^1} \;||\Psi||_{\G_+}\;\leq \frac{1}{\nu} \;e^{c_1 e^{c_1 t}} \;||\Psi||_{\G_+}^2\,.
\end{array}
\eean
Next, we show that there exists $C>0$ depending only on $\varphi_0$ such that
\bean
\left\|\hbarr^{-1} Q_s(z)^{Wick}R_\nu\right\|_{\L(\F_+^1,\F_-^1)}\leq C\,(\nu\,\hbarr^{1/2}+\nu^2\hbarr)\,.
\eean
The latter bound follows  by Cauchy-Schwarz inequality, Lemma \ref{main-est} and (\ref{energy-est}),
\bean
|\la \Phi,\frac{P_3(s)}{\hbarr}^{Wick} R_\nu\Psi\ra|&\leq& \sqrt{\hbarr}
||\varphi_t||_{L^\infty(\rz)} \left[\sum_{n=1}^{2\nu}
(n+1) ||\Phi^{(n)}||^2_{L^2(\rz^n)}\right]^{1/2} \;\left[\sum_{n=1}^{2\nu}
n^2 ||\Psi^{(n+1)}(x,x,.)||^2_{L^2(\rz^n)}\right]^{1/2}\\ \nm\ds
&+& \sqrt{\hbarr}
||\varphi_t||_{L^\infty(\rz)} \left[\sum_{n=1}^{2\nu}
(n+1) ||\Psi^{(n)}||^2_{L^2(\rz^n)}\right]^{1/2} \;\left[\sum_{n=1}^{2\nu}
n^2 ||\Phi^{(n+1)}(x,x,.)||^2_{L^2(\rz^n)}\right]^{1/2}\\ \nm\ds
&\leq& 2\nu \sqrt{\hbarr} ||\varphi_t||_{L^\infty(\rz)}\; 
||(\hbarr^{-1}N+1)^{1/2}\Phi||_\F\;||\Psi||_{\F_+^1}\\ \nm\ds
&+&2\nu \sqrt{\hbarr} ||\varphi_t||_{L^\infty(\rz)}\; 
||(\hbarr^{-1}N+1)^{1/2}\Psi||_\F\;||\Phi||_{\F_+^1}\,,
\eean
and a similar estimate for $P^{Wick}$,
\bean
|\la \Phi,P^{Wick} R_\nu\Psi\ra|\leq \nu^2 \hbarr^2\, ||\Phi||_{\F_+^1}\;||\Psi||_{\F_+^1}\,.
\eean
Hence we can check that
\bean
\int_0^t \left|\la \W(s)\Psi, \hbarr^{-1} Q_s(z)^{Wick} R_\nu U_2(s,0) \Psi\ra\right| \,ds &\leq& C (\nu\hbarr^{1/2}+\nu^2\hbarr)
\int_0^t ||\W(s)\Psi||_{\F_+^1} \;|| U_2(s,0) \Psi||_{\F_+^1} \;ds\,.
\eean
Now, by Lemma \ref{coh_lem3} and Proposition \ref{quad-propag}  we obtain
\bean
\ds\int_0^t \left\|\W(s)\Psi\right\|_{\F_+^1}
\; \left\| U_{2}(s,0)\Psi\right\|_{\F_+^1} ~ds&\leq &
\int_0^t e^{c_1 e^{c_1 s}}\left\|\Psi\right\|_{\G_+} \left\|U_{2}(s,0)\Psi\right\|_{\F_+^1}  ~ds \\
&\leq&    \int_0^t e^{c_2 e^{c_2 s}} \left\|\Psi\right\|_{\G_+} \,\left\|\Psi\right\|_{\F_+^1} ~ds \\
&\leq &   \; e^{ c e^{ c s}} \left\|\Psi\right\|_{\G_+}^2 \,.
\eean
A simple computation yields
\bean
A_2(s) R_\nu-R_\nu A_2(s)&=&\frac{1}{2} \left[\sigma(\frac{\hbarr^{-1} N+2}{\nu})-\sigma(\frac{\hbarr^{-1}N}{\nu})\right]
\left(\int_\rz \varphi_t(x)^2 \overline{z(x)}^2 \, dx\right)^{Wick} \\
&&+\frac{1}{2} \left[\sigma(\frac{\hbarr^{-1} N-2}{\nu})-\sigma(\frac{\hbarr^{-1}N}{\nu})\right]
\left(\int_\rz \overline{\varphi_t(x)}^2 z(x)^2 \, dx\right)^{Wick}\,.
\eean
We easily check  that
\bean
\left\|\sigma(\frac{\hbarr^{-1} N\pm 2}{\nu})-\sigma(\frac{\hbarr^{-1}N}{\nu})\right\|_{\L(\F_+^1)}\leq \frac{2}{\nu} ||\sigma'||_{L^\infty(\rz_+)}\,,
\eean
since $\hbarr^{-1}\d\Gamma(-\Delta)+\hbarr^{-1}N$ commute with $\hbarr^{-1} N$. Thus, using
(\ref{est.p2t}) there exists $c_0,c>0$ such that
\bean
\int_0^t \left|\la \W(s)\Psi, [A_2(s), R_\nu] U_2(s,0) \Psi\ra\right| \,ds &\leq& \frac{c_0}{\nu}
\int_0^t \left\| \W(s)\Psi\right\|_{\F_+^1} \;
\left\|U_2(s,0) \Psi\right\|_{\F_+^1}\;ds\\
&\leq& \frac{1}{\nu} e^{c e^{c t}} \left\|\Psi\right\|_{\G_+}^2\,.
\eean
Finally, the claimed inequality in Proposition \ref{coherent} follows by collecting the previous estimates and letting
$\nu=\hbarr^{-1/4}$. \hfill\cqfd

We have the following two corollaries.
\begin{cor}
\label{hepp}
For any $\varphi_0\in H^2(\rz)$ and any $\xi\in L^2(\rz)$ we have the strong limit
$$
\ds s-\lim_{\hbarr\to 0} W(\frac{\sqrt{2}}{i\hbarr}\varphi_0)^*
\, e^{it/\hbarr H_\hbarr} \,W(\xi)\, e^{-it/\hbarr H_\hbarr}\, W(\frac{\sqrt{2}}{i\hbarr}\varphi_0)=e^{i\sqrt{2}{\rm Re}\la\xi,\varphi_t\ra}\,1\,,
$$
where $\varphi_t$  solves the NLS equation (\ref{nls}) with initial data $\varphi_0$.
\end{cor}
\proof
It is enough to prove for any $\Psi,\Phi\in\G_+$ the limit:
\bea
\label{eq.2}
\lim_{\hbarr\to 0} \la e^{-it/\hbarr H_\hbarr} W(\frac{\sqrt{2}}{i\hbarr}\varphi_0)\,\Psi,\,W(\xi)\,
e^{-it/\hbarr H_\hbarr}\, W(\frac{\sqrt{2}}{i\hbarr}\varphi_0)\Phi\ra=e^{i\sqrt{2}{\rm Re}(\xi,\varphi_t)}\,\la \Psi,\Phi\ra\,.
\eea
Indeed, using Proposition \ref{coherent}, we show
\bean
\la e^{-it/\hbarr H_\hbarr} W(\frac{\sqrt{2}}{i\hbarr}\varphi_0)\Psi,W(\xi)
e^{-it/\hbarr H_\hbarr} W(\frac{\sqrt{2}}{i\hbarr}\varphi_0)\Phi\ra&=&
\la  W(\frac{\sqrt{2}}{i\hbarr}\varphi_t) U_2(t,0)\Psi,W(\xi)\, W(\frac{\sqrt{2}}{i\hbarr}\varphi_t)U_2(t,0)\Phi\ra \\
&+&O(\hbarr^{1/8}).
\eean
Therefore by Weyl commutation relations we have
\bean
\la  W(\frac{\sqrt{2}}{i\hbarr}\varphi_t)\,U_2(t,0)\Psi,\,W(\xi)\, W(\frac{\sqrt{2}}{i\hbarr}\varphi_t)\,U_2(t,0)\Phi\ra=
\la  U_2(t,0)\Psi,\,W(\xi)\, U_2(t,0)\Phi\ra e^{i\sqrt{2}{\rm Re}(\xi,\varphi_t)}\,,
\eean
Thus the limit is proved since $s-\lim_{\hbarr\to 0}W(\xi)=1$. \hfill\cqfd

\bigskip

Recall that  $\F_0$ is the subspace of $\F$ spanned by vectors $\Psi\in\F$ such that $\Psi^{(n)}=0$ for any index $n\in\nz$ except for
finite number. Note that $\F_0\cap\G_+$ is dense in $\F$.
\begin{cor}
\label{wick-propag}
For any $\varphi_0\in H^2(\rz)$ and any $\Psi,\Phi\in \F_0\cap\G_+$ and $b\in\P_{p,q}(L^2(\rz))$, we have
$$
\ds \lim_{\hbarr\to 0} \la W(\frac{\sqrt{2}}{i\hbarr}\varphi_0)\Psi, \, e^{it/\hbarr H_\hbarr} \,b^{Wick}\, e^{-it/\hbarr H_\hbarr}\,
W(\frac{\sqrt{2}}{i\hbarr}\varphi_0)\Phi\ra=b(\varphi_t)\,\la\Psi,\,\Phi\ra\,,
$$
where $\varphi_t$  solves the NLS equation (\ref{nls}) with initial data $\varphi_0$.
\end{cor}
\proof
Consider a $(p,q)$-homogenous polynomial $b\in\P_{p,q}(L^2(\rz))$.
We have
\bean
\mathcal{A}&:=&\la W(\frac{\sqrt{2}}{i\hbarr}\varphi_0)\Psi, \, e^{it/\hbarr H_\hbarr} \,b^{Wick}\, e^{-it/\hbarr H_\hbarr}\,
W(\frac{\sqrt{2}}{i\hbarr}\varphi_0)\Phi\ra\\
&=& \la (N+1)^{q}W(\frac{\sqrt{2}}{i\hbarr}\varphi_0)\Psi, e^{it/\hbarr H_\hbarr} \, B_\hbarr \,
e^{-it/\hbarr H_\hbarr} (N+1)^{p} W(\frac{\sqrt{2}}{i\hbarr}\varphi_0)\Phi\ra\,,
\eean
where $B_\hbarr:=(N+1)^{-q} b^{Wick} (N+1)^{-p}$. The number estimate (\ref{number_est}) yields
$$
\left\|B_\hbarr\right\|\leq \left\|\tilde{b}\right\|_{\L(L_s^2(\rz^p),L_s^2(\rz^q))}\,,
$$
uniformly in $\hbarr\in(0,1]$. Let $\tilde N_t$ be the positive operator given by
$$\tilde N_t=N+2{\rm Re}\la z,\varphi_t\ra^{Wick}+||\varphi_t||_{L^2(\rz)}^2\,.
$$
By (\ref{wick-weyl}),  we get
\bean
\mathcal{A}=\la W(\frac{\sqrt{2}}{i\hbarr}\varphi_0)(\tilde N_0+1)^{q}\Psi, \, e^{it/\hbarr H_\hbarr}
\,B_\hbarr \, e^{-it/\hbarr H_\hbarr}\,  W(\frac{\sqrt{2}}{i\hbarr}\varphi_0)(\tilde N_0+1)^{p}\Phi\ra\,.
\eean
Now, observe that
$$
\lim_{\hbarr\to 0} (\tilde N_0+1)^{p}\Phi=(1+||\varphi||_{L^2(\rz)}^2)^{p} \Phi \hspace{.2in}
\mbox{ and } \hspace{.2in} \lim_{\hbarr\to 0} (\tilde N_0+1)^{q}\Psi=(1+||\varphi||_{L^2(\rz)}^2)^{q} \Psi\,.
$$
So, using Proposition \ref{coherent} we obtain
\bean
\mathcal{A}&=& (1+||\varphi_0||_{L^2(\rz)}^2)^{p+q} \;
\la W(\frac{\sqrt{2}}{i\hbarr}\varphi_t)\, U_2(t,0)\Psi, \,  \,B_\hbarr\,
W(\frac{\sqrt{2}}{i\hbarr}\varphi_t) U_2(t,0)\Phi\ra+O(\hbarr^{1/8}) \\
&=& \la U_2(t,0)\Psi, \,  (\tilde N_t+1)^{-q}\,b(.+\varphi_t)^{Wick}\,(\tilde N_t+1)^{-p}\, U_2(t,0)\Phi\ra+O(\hbarr^{1/8})\,.
\eean
We set $\Psi_\hbarr=(N+1)^{q}(\tilde N_t+1)^{-q}U_2(t,0)\Psi$ and $\Phi_\hbarr=(N+1)^{p}(\tilde N_t+1)^{-p}U_2(t,0)\Phi$ and
remark that we can show for $\varphi_0\neq 0$ and $\mu$ a positive integer the following strong limit
\bea
\label{sconvnum}
s-\lim_{\hbarr\to 0}(N+1)^\mu (\tilde N_t+1)^{-\mu}=\frac{1}{(1+||\varphi_t||_{L^2(\rz)}^2)^\mu}\,.
\eea
This holds since we have by explicit computation
\bean
||(a(\varphi_t)+a^*(\varphi_t))(N+||\varphi_t||^2+1)^{-1}||\leq \frac{||\varphi_t||}{2\sqrt{||\varphi_t||^2+1}}+\frac{||\varphi_t||}{2
\sqrt{||\varphi_t||^2+1-\hbarr}}<1,
\eean
for $\hbarr$ sufficiently small and hence we can write
\bean
(N+1) (\tilde N_t+1)^{-1}=(N+1) (N+||\varphi_t||^2+1)^{-1} 
[\overbrace{(a(\varphi_t)+a^*(\varphi_t))(N+||\varphi_t||^2+1)^{-1}}^{\mathcal{R}_\hbarr}+1]^{-1}\,.
\eean
This proves (\ref{sconvnum}) for $\mu=1$ since $s-\lim_{\hbarr\to 0}\mathcal{R}_\hbarr=0$. 
Now, we proceed by induction on $\mu$ using a commutator argument
\bean
(N+1)^{\mu+1} (\tilde N_t+1)^{-(\mu+1)}&=&(N+1)^{\mu} (\tilde N_t+1)^{-\mu} (N+1) (\tilde N_t+1)^{-1}
\\& +&
(N+1)^\mu (\tilde N_t+1)^{-\mu} [(\tilde N_t+1)^\mu,N] (\tilde N_t+1)^{-(\mu+1)}\,,
\eean
with the observation that the second term of (r.h.s.) converges strongly to $0$.
Therefore, we obtain
$$
\lim_{\hbarr\to 0}\Psi_\hbarr= \frac{1}{(1+||\xi||_{L^2(\rz)}^2)^q}U_2(t,0)\Psi\quad \mbox{ and } \quad
\lim_{\hbarr\to 0}\Phi_\hbarr=\frac{1}{(1+||\xi||_{L^2(\rz)}^2)^p}U_2(t,0)\Phi\,.
$$
It is also easy to show by explicit computation that
\bean
w-\lim_{\hbarr\to 0} (N+1)^{-q} b_{r,s}^{Wick}( N+1)^{-p} =0\,,
\eean
for any $b_{r,s}\in\P_{r,s}(L^2(\rz))$ such that $0<r\leq p$ and $0<s\leq q$.
Hence, letting $\hbarr\to 0$, we get
\bean
\lim_{\hbarr\to 0} \mathcal{A} &=& (1+||\varphi_0||_{L^2(\rz)}^2)^{p+q} \;
\lim_{\hbarr\to 0} \la \Psi_\hbarr,  (N+1)^{-q} b(\varphi_t) \,( N+1)^{-p}
\Phi_\hbarr\ra \\
&=& b(\varphi_t) \, \la U_2(t,0)\Psi, \,U_2(t,0)\Phi\ra=b(\varphi_t)\;\la \Psi, \,\Phi\ra \,,
\eean
since $||\varphi_t||_{L^2(\rz)}= ||\varphi_0||_{L^2(\rz)}$ and $s-\lim_{\hbarr\to 0} (N+1)^{-\mu}=1$ for $\mu>0$.\hfill\cqfd

\bigskip

We identify the propagator $U_2(t,s)$ as a time-dependent Bogoliubov's transform on the
Fock representation of the Weyl commutation relations.

\begin{prop}
\label{ccr}
Let $\varphi_0\in H^2(\rz)$ and consider the propagator $U_2(t,0)$ given in Proposition \ref{quad-propag}. For a given $s\in\rz$
let $\xi_s\in H^2(\rz)$, we have
\bean
U_2(t,s) \,W(\frac{\xi_s}{i\sqrt{\hbarr}}) \, U_2(s,t)=W(\frac{\beta(t,s)\xi_s}{i\sqrt{\hbarr}})\,
\eean
where $\beta(t,s)$ is the symplectic propagator on $L^2(\rz)$, solving the equation
\bea
\label{sympl-eq}
\left\{
\begin{array}{l}
i\partial_t \xi_t(x)=[-\Delta+2|\varphi_t(x)|^2]\;\xi_t(x)+\varphi_t(x)^2 \;\overline{\xi_t(x)}\,,\\
\xi_{|t=s}=\xi_s\,,
\end{array}
\right.
\eea
such that $\beta(t,s)\xi_s=\xi_t$.
\end{prop}
\proof
Observe that if $\varphi_0\in H^2(\rz)$ then the solution $\varphi_t$ of the NLS equation
(\ref{nls}) with initial condition $\varphi_0$ satisfies  $\varphi_t\in C^0(\rz, L^\infty(\rz))$. Hence,
by standard arguments the equation (\ref{sympl-eq}) admits a unique solution  $\xi_t\in C^0(\rz,H^2(\rz))\cap
C^1(\rz, L^2(\rz))$ for any $\xi_s\in H^2(\rz)$. Moreover, the propagator
$$
\beta(t,s)\xi_s=\xi_t,
$$
defines a symplectic transform on $L^2(\rz)$ for any $t,s\in\rz$. This follows by differentiating
$$
{\rm Im}\la \beta(t,s)\xi,\beta(t,s)\eta\ra\,,
$$
with respect to $t$ for $\xi,\eta\in H^2(\rz)$. Furthermore, $\beta$ satisfies the laws
$$
\beta(s,s)=1, \hspace{.2in} \beta(t,s)\beta(s,r)=\beta(t,r)\, \hspace{.2in} \mbox{ for } \hspace{.2in} t,r,s\in\rz.
$$
Now, we differentiate with respect to $t$ the quantity
\bean
U_2(s,t)\,W(\frac{\xi_t}{i\sqrt{\hbarr}}) \, U_2(t,s)\,
\eean
in the sense of quadratic forms on $\F_+^1$, with $\xi_t$ solution of (\ref{sympl-eq}).
Hence, using Lemma \ref{diff-weyl} (ii), we get
\bea
\label{weyl_imp}
\begin{array}{lll}
\ds \partial_t \left[  U_2(s,t) \, W(\frac{\sqrt{2}}{i\sqrt{\hbarr}} \xi_t)\,U_2(t,s)\right]
&=& U_2(s,t) W(\frac{\sqrt{2}}{i\sqrt{\hbarr}} \xi_t)
\left[W(\frac{\sqrt{2}}{i\sqrt{\hbarr}} \xi_t)^* iA_2(t) W(\frac{\sqrt{2}}{i\sqrt{\hbarr}} \xi_t)-iA_2(t)\right.\\ \nm\ds
&&-i \left.\left(
{\rm Re}\la \xi_t,i\partial_t\xi_t\ra+\frac{2}{\sqrt{\hbarr}}
{\rm Re}\la z, i\partial_t\xi_t\ra^{Wick}\right) \right] U_2(t,s)\,.
\end{array}
\eea
Now, by \cite[Lemma 2.10]{AmNi1}, we obtain
\bean
W(\frac{\sqrt{2}}{i\sqrt{\hbarr}} \xi_t)^* A_2(t) W(\frac{\sqrt{2}}{i\sqrt{\hbarr}} \xi_t)&=& \hbarr^{-1} m(t)[z+\sqrt{\hbarr} \xi_t]^{Wick}\,,
\eean
where $m(t)[z]$ is the continuous polynomial on $\S(\rz)$ given by
\bean
m(t)[z]=\la z, -\Delta z\ra+P_2(t)[z]\,.
\eean
Therefore, the (r.h.s.) of (\ref{weyl_imp}) is null if we show  that
\bean
m(t)[z+\sqrt{\hbarr}\xi_t]-m(t)[z]-\left(\hbarr
{\rm Re}\la \xi_t,i\partial_t\xi_t\ra+2\sqrt{\hbarr}
{\rm Re}\la z, i\partial_t\xi_t\ra\right)=0\,.
\eean
This follows by straightforward computation.
\hfill\cqfd

\section{Propagation of chaos}
\label{sec.chaos}
Propagation of chaos for a many-boson system with  point pair-interaction in one dimension was studied  in
\cite{ABGT} (see also the related work \cite{AGT}). Here we prove this conservation  hypothesis for such quantum system
using the method in \cite{RodSch}. Thus, we are led to study the asymptotics of time-evolved Hermite states
\bean
e^{-it/\hbarr_n H_{\hbarr_n}} \,\varphi_0^{\otimes n} \quad\mbox{ with}\quad \varphi_0\in H^2(\rz),
\;||\varphi_0||_{L^2(\rz)}=1\,,
\eean
when  $n\to\infty$ with $n\hbarr_n=1$. We denote the coherent states by
$$
E(\varphi_0):=W(\frac{\sqrt{2}}{i\hbarr} \varphi_0)\Omega_0,
$$
where $\Omega_0=(1,0,\cdots)$ is the vacuum vector in the Fock space $\F$. To pass from coherent states
to Hermite states we use the integral representation proved in \cite{RodSch},
\bea
\label{rodsch}
\hspace{1in}\varphi_0^{\otimes n}=\frac{\gamma_n}{2\pi} \, \int_0^{2\pi} e^{-i \theta n}\; E(e^{i\theta}\varphi_0)\;d\theta\,,
\hspace{.2in} \mbox{ where }\hspace{.2in}\gamma_n:=\frac{e^{1/2\hbarr_n}\sqrt{n!}}{\hbarr_n^{-n/2} }.
\eea
Asymptotically, the factor $\gamma_n$ grows as $(2\pi n)^{1/4}$ when $n\to\infty$.

\bigskip

In the following proposition we prove the  chaos conservation hypothesis.
\begin{prop}
\label{chaos}
For any $\varphi_0\in H^2(\rz)$ such that $||\varphi_0||_{L^2(\rz)}=1$ and any $b\in\P_{p,p}(L^2(\rz))$, we have
\bean
\lim_{n\to \infty}\la \varphi_0^{\otimes n}, e^{it/\hbarr_n H_{\hbarr_n}} \; b^{Wick} \;
e^{-it/\hbarr_n H_{\hbarr_n}}\varphi_0^{\otimes n}\ra=b(\varphi_t)\,,
\eean
where $n\hbarr_n=1$ and $\varphi_t$ solves the NLS equation (\ref{nls}) with initial data $\varphi_0$.
\end{prop}
\proof
It is  known that if a sequence of positive trace-class operators $\rho_n$ on $L^2(\rz)$ converges in the
weak operator topology to $\rho$ such that $\lim_{n\to\infty}{\rm Tr}[\rho_n]={\rm Tr}[\rho]<\infty$ then $\rho_n$
converges in the trace norm to $\rho$ (see, for instance \cite{DA}). This argument reduces the proof to the case
$$
b(z)=\prod_{i=1}^p  \la z,f_i\ra\;\la g_i,z\ra\,,
$$
where $f_i,g_i\in L^2(\rz)$. For shortness, we set
$$
E_\theta=E(e^{i\theta}\varphi_0) \quad\mbox{ and } \quad E_\theta^t=
e^{-it/\hbarr_n H_{\hbarr_n}}E_\theta\,.
$$
Using formula (\ref{rodsch}), we get
\bean
\Gamma_n:=\la \varphi_0^{\otimes n},e^{it/\hbarr_n H_{\hbarr_n}}\,
b^{Wick} \,e^{-it/\hbarr_n H_{\hbarr_n}}\varphi_0^{\otimes n}\ra =  \frac{\gamma_n^2}{(2\pi)^2}\int_{[0,2\pi]^2}\;  e^{-in(\theta-\theta')}
\la  E_{\theta'}^t,\, b^{Wick}\, E_\theta^t\ra\;d\theta d\theta'\,.
\eean
It is easily seen that
$$
(N+1)^{-p} e^{-it/\hbarr_n H_{\hbarr_n}}\varphi_0^{\otimes n}=2^{-p}e^{-it/\hbarr_n H_{\hbarr_n}}\varphi_0^{\otimes n}\,.
$$
Therefore, we write
\bean
\Gamma_n= \frac{4^p\gamma_n^2}{(2\pi)^2}\int_{[0,2\pi]^2}\;  e^{-in(\theta-\theta')}
\la  E_{\theta'}^t,\,
(N+1)^{-p}\prod_{i=1}^p a^*(f_i)\prod_{j=1}^p a(g_j) (N+1)^{-p}\, E_\theta^t\ra\;d\theta d\theta'\,.
\eean
Now, we use the decomposition
\bean
\ds \prod_{i=1}^p a^*(f_i)\prod_{j=1}^p a(g_j)&=& \ds\sum_{I,J\subset \mathcal{N}_p} \;\prod_{i\in I^c} [a^*(f_i)-\la \varphi_t^{\theta'},f_i\ra ]
\; \prod_{j\in J^c} [a(g_j)-\la g_j,\varphi_t^\theta\ra]  e^{-i (\#I\theta'-\#J\theta)} \\ \nm\ds
&\times& \prod_{i\in I} \overline{\la f_i,\varphi_t\ra}\; \prod_{j\in J} \la g_j,\varphi_t\ra\,,
\eean
where the sum runs over all subsets $I,J$ of $\mathcal{N}_p :=\{1,\cdots,p\}$. Thus, we can write
\bea
\label{sum}
\ds\Gamma_n-b(\varphi_t)=\ds
\sum_{I,J\subset\mathcal{N}_p}^{\#I+\#J<2p} \frac{4^p\gamma_n^2}{(2\pi)^2} \int_{[0,2\pi]^2} \;
e^{-i [(n-\#J)\theta- (n-\#I)\theta']} \la  \tilde E_{\theta'}^t, \,B_{I,J}^{Wick}\,
\tilde E_{\theta}^t\ra\;d\theta d\theta' \,,
\eea
where $\tilde E_{\theta}^t:=(N+1)^{-p}\, E_\theta^t$ and  $B_{I,J}(z)$ are sums of homogenous polynomials such that
\bean
\la \tilde E_{\theta'}^t,
B^{Wick}_{I,J} \tilde E_{\theta}^t\ra =\ds
\prod_{i\in I} \la \varphi_t, f_i\ra \; \prod_{j\in J} \la g_j,\varphi_t\ra
\times\left\langle \prod_{i\in I^c} [a(f_i)-\la f_i,\varphi_t^{\theta'}\ra] \tilde E_{\theta'}^t,
\prod_{j\in J^c} [a(g_{j})-\la g_j,\varphi_t^\theta\ra] \tilde E_{\theta}^t\right\rangle\,.
\eean
We have, for $0\leq \#I,\#J<p$, by Cauchy-Schwarz inequality
\bean
&&\hspace{-.6in}\left|\la \tilde E_{\theta'}^t,
B^{Wick}_{I,J} \tilde E_{\theta}^t\ra \right|\leq \,\prod_{i\in I,j\in J} ||g_j||_{L^2(\rz)} \;||f_i||_{L^2(\rz)}  \\ \nm\ds &&\times\;\;
\left\|\prod_{i\in I^c} [a(f_i)-\la f_i, \varphi_t^{\theta'}\ra] \tilde E_{\theta'}^t\right\|_{\F} \times \;\;
\left\|\prod_{j\in J^c} [a(g_j)-\la g_j,\varphi_t^\theta\ra] \tilde E_{\theta}^t\right\|_{\F}\,.
\eean
In the following we make use of the positive self-adjoint operator
\bean
\tilde N:=N+2{\rm Re}\la z,\varphi_t\ra^{Wick}+||\varphi_t||^2 1\,.
\eean
Observe that we have for any $\theta'\in [0,2\pi]$ and  $r\geq 1$,
\bean
\left\|\prod_{i=1}^r [a(f_i)-\la f_i,\varphi_t^{\theta'}\ra] \tilde E_{\theta'}^t\right\|_{\F}
&=& \left\|\prod_{i=1}^r a(f_i) (\tilde N+1)^{-p} \W(t)\Omega_0\right\|_{\F}\,\\
&\leq&\left\| \prod_{i=1}^{r-1} a(f_i) (\tilde N+1)^{-p} a(f_r) \W(t)\Omega_0\right\|_\F \\
&+&\left\|
\prod_{i=1}^{r-1} a(f_i) [a(f_r),(\tilde N+1)^{-p}] \W(t)\Omega_0\right\|_\F\,.
\eean
We easily show that
\bean
\left\|a(f_r) \W(t)\Omega_0\right\|_\F &\leq & ||f_r||_{L^2(\rz)}\; \sqrt{\hbarr_n} \,
\left\|\W(t)\right\|_{\L(\G_+,\F_+^1)}\,.
\eean
Furthermore, we have  
\bean
\left\| [a(f_r),(\tilde N+1)^p] (\tilde N+1)^{-p}\right\|_{\L(\F)}\leq C \;\hbarr_n\,,
\eean
using (\ref{numberequiv}) and the fact that $[a(f_r),(\tilde N+1)^p]$ is a Wick polynomial where we gained $\hbarr_n$ in its 
symbol, see \cite[Proposition 2.7 (ii)]{AmNi1}.
Recall also that we have by the number estimate (\ref{number_est}) and (\ref{numberequiv}),
\bean
\left\| \prod_{i=1}^{r-1} a(f_i) (\tilde N+1)^{-p}\right\|_{\L(\F)}\leq  C\,,
\eean
uniformly in $n$ and $\theta'\in[0,2\pi]$. Therefore, we have
\bea
\label{cha-est}
\left|\sum_{0\leq \#I,\#J<p} \frac{4^p\gamma_n^2}{(2\pi)^2} \int_{[0,2\pi]^2} \;
e^{-i [(n-\#J)\theta- (n-\#I)\theta']}
\la  \tilde E_{\theta'}^t, \,B_{I,J}^{Wick}\,
\tilde E_{\theta}^t\ra\;d\theta d\theta' \right|\leq C \;\gamma_n^2 \hbarr_n^{2p-(\#I+\#J)}\stackrel{n\rightarrow \infty}{\longrightarrow 0}\,.
\eea
It still to control the terms $\#I=p,\#J=p-1$ and $\#I=p-1,\#J=p$ which are similar. In fact, remark that we have
\bean
&&\hspace{-.8in}\frac{4^p\gamma_n^2}{(2\pi)^2} \int_{[0,2\pi]^2} \;
e^{-i [(n-p)\theta- (n-p+1)\theta']}
\la  \tilde E_{\theta'}^t, \,B_{I,\mathcal{N}_p}^{Wick}\,
\tilde E_{\theta}^t\ra\;d\theta d\theta' =\\ &&\hspace{.6in}\frac{4^p\gamma_n}{2\pi} \int_0^{2\pi} e^{i (n-p+1)\theta'}
\la  \tilde E_{\theta'}^t, \,B_{I,\mathcal{N}_p}^{Wick}\,\;
e^{it/\hbarr_n H_{\hbarr_n}}\;\varphi_0^{\otimes (n-p)}\ra\; d\theta'\,.
\eean
Now, a similar estimate as (\ref{cha-est}) yields that
\bean
\left|
\frac{4^p\gamma_n^2}{(2\pi)^2} \int_{[0,2\pi]^2} \;
e^{-i [(n-p)\theta- (n-p+1)\theta']} \;\;
\la  E_{\theta'}^t, \,B_{I,\mathcal{N}_p}^{Wick}\,
\tilde E_{\theta}^t\ra\;\;d\theta d\theta'\right|\leq C \;\gamma_n \sqrt{\hbarr_n}\stackrel{n\rightarrow \infty}{\longrightarrow 0}\,.
\eean
Thus, we conclude that $\ds\lim_{n\to\infty}\Gamma_n-b(\varphi_t)=0$.\hfill\cqfd

\begin{remark}\ \\
{\bf 1)} Let $\gamma_{k,n}^t$ be the $k$-particle correlation functions, defined by (\ref{corldef}),  associated to
the states $e^{-it/\hbarr_n H_{\hbarr_n}}\varphi_0^{\otimes n}$. Then Proposition \ref{chaos}
implies the following convergence in the trace norm
\bean
\lim_{n\to\infty}\gamma_{k,n}^t=\varphi_t(x_1)\cdots\varphi_t(x_k)\;\overline{\varphi_t(y_1)\cdots\varphi_t(y_k)}\,.
\eean
{\bf 2)} In terms of Wigner measures, introduced in \cite{AmNi1,AmNi2}, Proposition \ref{chaos} says that
the sequence $(e^{-it/\hbarr_n H_{\hbarr_n}}\varphi_0^{\otimes n})_{n\in\nz}$ admits a unique  (Borel probability)
Wigner measure $\mu_t$ given by
\bean
\mu_t=\frac{1}{2\pi}\int_0^{2\pi} \,\delta_{e^{i\theta} \varphi_t} \; d\theta\,,
\eean
where $\delta_{e^{i\theta} \varphi_t}$ is the Dirac measure on $L^2(\rz)$ at the point $e^{i\theta} \varphi_t$.
\end{remark}

\bigskip

\appendix
\begin{center}
{\bf\Large Appendix}
\end{center}
\section{Elementary estimate}
\begin{lem}
\label{main-est}
For any $\alpha>0$ and any $\Psi^{(n)}\in\S_s(\rz^n)$, we have
\bea
\label{est.A1}
\int_{\rz^{n-1}}|\Psi^{(n)}(x_2,x_2,\cdots,x_n)|^2 dx_2\cdots dx_n \leq \frac{\alpha}{\sqrt{2}} \la D_{x_1}^2
\Psi^{(n)}, \Psi^{(n)}\ra_{L^2(\rz^n)}+ \frac{\alpha^{-1}}{2\sqrt{2}}|\Psi^{(n)}|^2_{L^2(\rz^n)}\,.
\eea
\end{lem}
\proof
Let $x',\xi'\in \rz^{n-1}$ and $g\in\S(\rz^n)$. Let us denote the Fourier transform of $g$ by
$$
\hat g(\xi)=\int_{\rz^n} e^{-i x\xi} g(x) \,dx.
$$
We have
$$
g(0,x')=\frac{1}{(2\pi)^{n-1}} \int_{\rz^{n-1}} e^{i x' \xi'} \left(
\frac{1}{2\pi} \int_{\rz} \hat g(\xi_1,\xi') \,d\xi_1 \right) \,d\xi'\,.
$$
Cauchy-Schwarz inequality yields
\bean
\left|\int_\rz \hat g(\xi_1,\xi') \,d\xi_1 \right|^2\leq \int_\rz \left|\hat g(\xi_1,\xi')\right|^2
(\alpha^{-1}+\alpha \xi_1^2) \,d\xi_1 \times \int_\rz \frac{d\xi_1}{\alpha^{-1}+\alpha \xi_1^2}\,.
\eean
Therefore, we get
\bean
\int_{\rz^{n-1}} \left|g(0,x')\right|^2 \,dx'&=& \frac{1}{4\pi^2 (2\pi)^{n-1}} \int_{\rz^{n-1}}
\left|\int_\rz \hat g(\xi_1,\xi') \,d\xi_1 \right|^2 \,d\xi' \\
&\leq & \frac{1}{2(2\pi)^{n}}\int_{\rz^{n}}|\hat g(\xi_1,\xi')|^2 \,(\alpha^{-1}+\alpha \xi_1^2) \,d\xi_1 d\xi' \,.
\eean
Set $g(x_1,\cdots,x_n)=\Psi^{(n)}(\frac{x_1+x_2}{\sqrt{2}},\frac{x_2-x_1}{\sqrt{2}},x_3,\cdots,x_n)$, we obtain
\bean
\int_{\rz^{n-1}} \left|\Psi^{(n)}(x_2,x_2,\cdots,x_n)\right|^2 dx_2\cdots dx_n
&=& \frac{1}{\sqrt{2}} \int_{\rz^{n-1}} \left|g^{(n)}(0,x_2,\cdots,x_n)\right|^2 dx_2\cdots dx_n \\
&\leq& \frac{(2\pi)^{-n}}{2\sqrt{2}} \int_{\rz^{n}}|\hat g^{(n)}(\xi_1,\xi')|^2 \,(\alpha^{-1}+\alpha \xi_1^2+\alpha \xi_2^2) \,d\xi_1 d\xi'\\
&\leq& \frac{(2\pi)^{-n}}{2\sqrt{2}} \int_{\rz^{n}}|\hat \Psi^{(n)}(\xi_1,\xi')|^2 \,(\alpha^{-1}+\alpha \xi_1^2+\alpha \xi_2^2) \,d\xi_1 d\xi'\,.
\eean
Thus, by Plancherel's identity we obtain
$$
\int_{\rz^{n-1}}|\Psi^{(n)}(x_2,x_2,\cdots,x_n)|^2 dx_2\cdots dx_n \leq \frac{\alpha}{2\sqrt{2}} \la (D_{x_1}^2+D_{x_2}^2)
\Psi^{(n)}, \Psi^{(n)}\ra_{L^2(\rz^n)}+ \frac{\alpha^{-1}}{2\sqrt{2}}|\Psi^{(n)}|^2_{L^2(\rz^n)}\,.
$$
Thanks to the symmetry of $\Psi^{(n)}$, it is easy to see that
$$
\la (D_{x_1}^2+D_{x_2}^2)\Psi^{(n)},\Psi^{(n)}\ra=2 \la D_{x_1}^2\Psi^{(n)},\Psi^{(n)}\ra\,.
$$
Hence, we arrive at the claimed  estimate (\ref{est.A1}).  \hfill\cqfd

\section{Commutator theorems}
Here we first recall  an abstract regularity argument from Faris-Lavine work \cite[Theorem 2]{FL}.
\begin{thm}
\label{farislavine}
Let $A$ be a self-adjoint  operator and let $S$ be a positive self-adjoint
operator satisfying
\begin{itemize}
 \item $\D(S)\subset \D(A)$,
 \item $\pm i \left[\la A\Psi, S\Psi\ra-\la S\Psi, A\Psi\ra\right]\leq c ||S^{1/2}\Psi||^2$ for all $\Psi\in \D(S)$.
\end{itemize}
Then $\mathcal{Q}(S)$ is invariant by $e^{-it A}$ for any $t\in\rz$ and the inequality
\bean
||S^{1/2}e^{-it A} \Psi||\leq e^{c |t|} \;||S^{1/2}\Psi||\,
\eean
holds true.
\end{thm}
Next we recall the Nelson commutator theorem (see, {\it e.g.}, \cite[Theorem X.36']{RS},\cite{N}) with a useful regularity property
added as a consequence of Faris-Lavine's Theorem \ref{farislavine}.
\begin{thm}
\label{nelson}
Let $S$ be a self-adjoint operator on a Hilbert space $\H$ such that $S\geq 1$. Consider
a quadratic form $a(.,.)$ with $\mathcal{Q}(a)=\D(S^{1/2})$ and satisfying:
\begin{description}
 \item[](i) $|a(\Psi,\Phi)|\leq c_1 ||S^{1/2} \Psi||\; ||S^{1/2} \Phi||$ for any $\Psi,\Phi\in \D(S^{1/2})$;
 \item[] (ii) $|a(\Psi,S\Phi)-a(S\Psi,\Phi)|\leq c_2 ||S^{1/2} \Psi||\; ||S^{1/2} \Phi||$ for any $\Psi,\Phi\in\D(S^{3/2})$.
\end{description}
Then the linear operator $A:\D(A)\to \H$, $\D(A)=\{\Phi\in\D(S^{1/2}): \H\ni\Psi\mapsto a(\Psi,\Phi)\mbox{ continuous }\}$
associated to the quadratic form $a(.,.)$  through the relation
\bean
\la \Psi,A\Phi\ra_\H=a(\Psi,\Phi) \mbox{ for all } \Psi\in\D(S^{1/2}),\Phi\in\D(A)
\eean
is densely defined and satisfies:
\begin{enumerate}
 \item $\D(S)\subset\D(A)$ and $||A\Psi||\leq c ||S\Psi||$ for any $\Psi\in \D(S)$;
 \item $A$ is essentially self-adjoint on any core of $S$;
 \item $e^{-it \tilde A}$ preserves $\D(S^{1/2})$ with the inequality
\bean
||S^{1/2}e^{-it \tilde A} \Psi||\leq e^{c_2 |t|} \;||S^{1/2}\Psi||\,
\eean
where $\tilde A$ denotes the self-adjoint extension of $A$.
\end{enumerate}
\end{thm}
\proof
The point (3) follows from Theorem \ref{farislavine} since its assumptions:
\begin{itemize}
 \item $\D(S)\subset\D(A)$,
 \item $\pm i \left[\la A\Psi,S\Psi\ra-\la S\Psi,A\Psi\ra\right]\leq c_2 ||S^{1/2}\Psi||^2$, for any $\Psi\in\D(S)$,
\end{itemize}
hold true using items 1), 2) and hypothesis (ii).\hfill\cqfd

We naturally associate to a self-adjoint operator $S\geq 1$ acting on a Hilbert space $\H$, a Hilbert rigging
$\H_{\pm 1}$ where $\H_{+ 1}$ is defined as $\D(S^{1/2})$ endowed with the inner product
\bean
\la \psi,\phi\ra_{\H_{+1}}:=\la S^{1/2}\psi,S^{1/2}\phi\ra_{\H}\,,
\eean
and $\H_{-1}$ is the completion of $\D(S^{-1/2})$ with respect to the  inner product
\bean
\la \psi,\phi\ra_{\H_{-1}}:=\la S^{-1/2}\psi,S^{-1/2}\phi\ra_{\H}\,.
\eean
Assumption (ii) of Theorem \ref{nelson} can be reformulated in some other slightly different ways.

\begin{lem}
\label{equi-assmp-nels}
Consider a  self-adjoint operator $S$ satisfying $S\geq 1$ with the associated Hilbert rigging $\H_{\pm 1}$ defined above.
Let $A$ be a symmetric bounded operator in $\L(\H_{+1},\H_{-1})$, then the three following statements are equivalent,
\begin{description}
 \item[(1)] There exists $c>0$ such that for any $\Psi,\Phi\in\D(S^{3/2})$,
 $$|\la S\Psi,A\Phi\ra-\la A\Psi,S\Phi\ra|\leq c \;||\Psi||_{\H_{+1}}\; ||\Phi||_{\H_{+1}}\,,$$
 \item[(2)]  There exists $c>0$ such that for any $\Psi,\Phi\in\D(S^{1/2})$ and $\lambda>0$,
 $$|\la (\lambda S+1)^{-1}S\Psi,A(\lambda S+1)^{-1}\Phi\ra-\la A(\lambda S+1)^{-1}\Psi,(\lambda S+1)^{-1}S\Phi\ra|\leq
 c \;||\Psi||_{\H_{+1}}\; ||\Phi||_{\H_{+1}}\,,$$
 \item[(3)] There exists $c>0$ such that for any $\Psi,\Phi\in\D(S^{1/2})$  and $\lambda>0$,
 $$|\la (\lambda S+1)^{-1}S\Psi,A\Phi\ra-\la A \Psi,(\lambda S+1)^{-1}S\Phi)|\leq
 c \;||\Psi||_{\H_{+1}}\; ||\Phi||_{\H_{+1}}\,.$$
\end{description}
\end{lem}
\proof
$\bullet$  (1)$\Leftrightarrow$(2):\\
Observe that if $\lambda>0$ then $(\lambda S+1)^{-1}\D(S^{1/2})\subset \D(S^{3/2})$. Assume (1) and let us prove (2) for
$\Psi,\Phi\in\D(S^{1/2})$. Using (1) with $\tilde \Psi=(\lambda S+1)^{-1}\Psi\in\D(S^{3/2})$ and $\tilde \Phi=
(\lambda S+1)^{-1}\Phi\in\D(S^{3/2})$, we obtain
\bea
\label{commu2}
\left|\la S\tilde \Psi,A\tilde \Phi\ra-\la A\tilde \Psi,S\tilde\Phi\ra\right|\leq c \left\|(\lambda S+1)^{-1}\Psi\right\|_{\H_{+1}} \times
\left\|(\lambda S+1)^{-1}\Phi\right\|_{\H_{+1}}\,.
\eea
It is easy to see that the right hand side of (\ref{commu2}) is bounded by $c ||\Psi||_{\H_{+1}} ||\Phi||_{\H_{+1}}$.
Thus, we obtain (2). Now, to prove (2)$\Rightarrow$(1), we observe that $(\lambda S+1)\D(S^{3/2})\subset \D(S^{1/2})$ and use
(2) with $\Psi_\lambda=(\lambda S+1)\Psi\in\D(S^{1/2})$, $\Phi_\lambda=(\lambda S+1)\Phi\in\D(S^{1/2})$ such that
$\Psi,\Phi\in\D(S^{3/2})$. Therefore, we get for $\lambda>0$
\bea
\label{commu3}
\left|\la S \Psi,A \Phi\ra-\la A \Psi,S \Phi\ra\right|\leq c \left\|\Psi_\lambda\right\|_{\H_{+1}} \times
\left\|\Phi_\lambda\right\|_{\H_{+1}}\,.
\eea
Letting $\lambda\to 0$ in the right hand side of (\ref{commu3}), we obtain (2).

\noindent
$\bullet$ (2)$\Leftrightarrow$(3):\\
Let $\Psi,\Phi\in\D(S^{1/2})$ and $\lambda>0$, we have as identity in $\L(\H_{+1},\H_{-1})$
\bean
A (\lambda S+1) (\lambda S+1)^{-1}=A \lambda S(\lambda S+1)^{-1}+A (\lambda S+1)^{-1}\,,
\eean
since $\lambda S(\lambda S+1)^{-1}\in\L(\H_{+1})$ and $(\lambda S+1)^{-1}\in\L(\H_{+1})$. Therefore,
since $(\lambda S+1)^{-1}S\Psi\in\H_{+1}$ and $(\lambda S+1)^{-1}S\Phi\in\H_{+1}$, the following computation is justified
\bean
&&\hspace{-.2in}\la (\lambda S+1)^{-1}S\Psi,A\Phi\ra-\la A\Psi,(\lambda S+1)^{-1}S\Phi\ra
\\&&\hspace{.5in}=\la (\lambda S+1)^{-1}S\Psi,A(\lambda S+1)(\lambda S+1)^{-1}\Phi\ra-\la A(\lambda S+1)
(\lambda S+1)^{-1}\Psi,(\lambda S+1)^{-1}S\Phi\ra
\\&&\hspace{.5in}=\la (\lambda S+1)^{-1}S\Psi,A(\lambda S+1)^{-1}\Phi\ra-\la A(\lambda S+1)^{-1}\Psi,(\lambda S+1)^{-1}S\Phi\ra\,.
\eean
So, this shows the equivalence of the statements (2) and (3). \hfill\cqfd

\section{Non-autonomous Schr\"odinger equation}
\label{appendix_sch}
Consider the Hilbert rigging
$$
\H_{+}\subset \H\subset \H_{-}\,.
$$
This  means that $\H$ is a Hilbert space with an inner product $(.,.)_{\H}$ and $\H_{+}$ is a dense subspace of $\H$ which is itself a
Hilbert space with respect to another inner product $(.,.)_{\H_{+}}$ such that
$$
||u||_{\H}:=\sqrt{(u,u)_{\H}}\leq ||u||_{\H_{+}}:=\sqrt{(u,u)_{\H_{+}}}\hspace{.3in}\forall u\in\H_+\,.
$$
The Hilbert space $\H_-$ is  defined as the completion of $\H$ with respect to the norm
\bea
\label{rigged1}
||u||_{\H_-}:=\sup_{f\in\H_+, ||f||_{\H_+}=1}|(f,u)_{\H}|\,.
\eea
This extends by continuity the inner product $(.,.)_{\H}$ to a sesquilinear form on $\H_-\times \H_+$ satisfying
\bean
\left|(u,\xi)_{\H}\right|\leq  ||u||_{\H_+}\;||\xi||_{\H_-}\quad \forall u\in\H_+,\forall\xi\in\H_-\,.
\eean
Furthermore, we have
\bea
\label{rigged2}
||u||_{\H_+}=\sup_{\xi\in\H_-, ||\xi||_{\H_-}=1}|(u,\xi)_{\H}|\,.
\eea

Let $I$ be a closed interval of $\rz$ and let  $\big(A(t)\big)_{t\in I}$ denote a  family of self-adjoint operators  on $\H$
such that $\D(A(t))\cap\H_+$ is dense in $\H_+$ and $A(t)$ are continuously extendable to bounded operators in $\L(\H_+,\H_-)$. We aim to solve the following abstract non-autonomous Schr\"odinger equation
\bea
\label{abs_schrod}
\left\{
 \begin{array}[c]{l}
   i\partial_t u=A(t) u\,,\quad t\in I\\
   u(t=0)=u_0\,,
 \end{array}
\right.
\eea
where $u_0\in\H_+$ is given and $t\mapsto u(t)\in\H_+$ is the unknown. This  is a particular case of the more general topic
of solving non-autonomous Cauchy problems where $-iA(t)$ are infinitesimal  generators of  $C_0$-semigroups (see \cite{Si},\cite{Ki}).
We provide here a useful result  (Theorem \ref{thm_schrod}) which follows from the work of Kato \cite{Ka}.

\begin{defn}
\label{unit-propag}
We say that the map
$$
I\times I \ni (t,s)\mapsto U(t,s)
$$
is a unitary propagator of the problem (\ref{abs_schrod}) iff:\\
(a) $U(t,s)$ is unitary on $\H$,\\
(b) $U(t,t)=1$ and  $U(t,s) U(s,r)=U(t,r)$ for all $t,s,r\in I$,\\
(c) The map  $t\in I\mapsto U(t,s)$ belongs to $C^0(I,\L(\H_+))\cap C^{1}(I,\L(\H_+,\H_-))$
and satisfies
$$
i \partial_t U(t,s) \psi=A(t) U(t,s) \psi, \quad \forall \psi\in\H_+, \forall  t,s\in I.
$$
\end{defn}
Here  $C^k(I,\mathfrak{B})$ denotes the space of $k$-continuously differentiable $\mathfrak{B}$-valued functions where
$\mathfrak{B}$ is endowed with the strong operator topology.
\bigskip

\begin{thm}
\label{thm_schrod}
Let $I$ be a compact interval and let $\H_{+}\subset \H\subset \H_{-}$ be a Hilbert rigging with $\big(A(t)\big)_{t\in I}$
a family of self-adjoint operators  on $\H$ as above satisfying:\\
(i) $I\ni t\mapsto A(t)\in\L(\H_{+},\H_{-})$ is norm continuous.\\
(ii) $\rz\ni\tau\mapsto e^{i \tau A(t)}\in \L(\H_+)\,$ is strongly continuous.\\
(iii) There exists a family of Hilbertian norms $\big(||.||_t\big)_{t\in I}$ on $\H_+$ equivalent to $||.||_{\H_+}$ such that:
$$
\exists c>0, \forall \psi\in\H_+:\;\;\; ||\psi||_t\leq e^{c |t-s|} \, ||\psi||_s \;\; \mbox{ and }
\;\; ||e^{i \tau A(t)} \psi||_{t} \leq e^{c |\tau|} ||\psi||_{t}\,.
$$
Then  the non-autonomous Cauchy problem (\ref{abs_schrod}) admits a unique unitary propagator $U(t,s)$.\\
Moreover, the following estimate holds
$$
\forall \psi\in\H_{+},\quad ||U(t,s)\psi||_{t} \leq e^{2c |t-s|}\, ||\psi||_{s}\,.
$$
\end{thm}
\proof
We follow the same strategy as in \cite{Ka} and split the proof into three steps.
We assume, for reading convenience, that the interval $I$ is of the form $[0,T], T>0$
however the proof works exactly in the same way for any compact interval. Remark also that 
there is no restriction if we assume that $||.||_{\H_+}=||.||_0\,$.

\noindent
\underline{\bf Propagator approximation:}\\
Let $ (t_0,\cdots,t_n)$ be a regular partition of the interval $I$
with
$$
t_j=\frac{jT}{n},\;\; j=0,\cdots,n.
$$
Consider the sequence of operator-valued step functions defined by
$$
A_n(t):=A(T) 1_{\{T\}}(t)+\sum_{j=0}^{n-1} A(t_j) \, 1_{[t_j,t_{j+1}[}(t)\,,
$$
for any $n\in\nz^*$ and $t\in I$. Assumption (i) ensures that
$$
\lim_{n\to\infty}||A_n(t)-A(t)||_{\L(\H_+,\H_-)}=0\,,
$$
uniformly in $t\in I$. We now construct an approximating unitary propagator $U_n(t,s)$ as follows:
\bea
\label{propag}
\left\{
\begin{array}[c]{l}
\ds
\mbox{ - if } t_j\leq t,s\leq t_{j+1} \mbox{ then } U_n(t,s)=e^{-i(t-s) A(t_j)}\,
\\ \nm\ds
\mbox{ - if } t_j<s\leq t_{j+1}<\cdots <t_l\leq t<t_{l+1}  \mbox{ then } U_n(t,s)=e^{-i(t-t_l) A(t_l)}\cdots  e^{-i(t_{j+1}-s) A(t_j)}
\\\nm\ds
\mbox{ - if } t_j<t\leq t_{j+1}<\cdots <t_l\leq s<t_{l+1} \mbox{ then } U_n(t,s)=e^{-i(t-t_{j+1}) A(t_j)}\cdots  e^{-i(t_{l}-s) A(t_l)}\,,
\end{array}
\right.
\eea
for any $j=0,\cdots,n-1$ and $l=1,\cdots,n$ with $j<l$.\\
By definition, the operators $U_n(t,s)$ are unitary on $\H$ for $t,s\in I$  and satisfy
\bea
\label{gr_law1}
U_n(t,t)=1, \hspace{.2in} U_n(t,s)^*=U_n(s,t)\,.
\eea
Moreover, one can first check that
$$
U_n(t,s) U_n(s,r)=U_n(t,r) \mbox{ for } r\leq s\leq t, \mbox{ with } t,s,r\in I
$$
and then extend it for any $(t,s,r)\in I^3$ with the help of (\ref{gr_law1}).
Therefore, $U_n(t,s)$ satisfy the properties (a)-(b) of Definition \ref{unit-propag}. Again by (\ref{propag}) and assumptions (i)-(ii) we have
\bea
\label{schrodapprox}
i\partial_t U_n(t,s)\psi=A_n(t) U_n(t,s)\psi \hspace{.2in} \mbox{ and } \hspace{.2in} -i\partial_s U_n(t,s)\psi=U_n(t,s)A_n(s)\psi,
\eea
for any $\psi\in\H_+$ and any $t,s\neq t_j$, $j=0,\cdots,n$.

\bigskip
\noindent
\underline{\bf Convergence of the approximation:}\\
Assumption (iii) implies that
\bean
||e^{-i s_n A(t_n)} \cdots  e^{-i s_1 A(t_1)}\psi||_T\leq e^{cT} e^{ c (s_1+\cdots+s_n)} ||\psi||_0\,,
\eean
and
\bean
||e^{-i s_1 A(t_1)} \cdots  e^{-i s_n A(t_n)}\psi||_0\leq e^{cT} e^{ c (s_1+\cdots+s_n)} ||\psi||_T\,,
\eean
for any $s_j\geq 0$, $j=1,\cdots, n$. Hence, using the equivalence of the norms $||.||_0=||.||_{\H_+}$ and 
$||.||_T$ one shows the existence of $M>0$ ($M=e^{2cT}$) such that
\bea
\label{kat-est}
||U_n(t,s)||_{\L(\H_+)}\leq M \; e^{c |t-s|}\;\;\; \mbox{ and by duality } \;\;\;
||U_n(t,s)||_{\L(\H_-)}\leq M \; e^{c |t-s|}\,.
\eea
Furthermore, the same argument above yields
\bea
\label{gen-propag-est}
||U_n(t,s)\psi||_{t}\leq  e^{2 c (|t-s|+T/n)} ||\psi||_s.
\eea
Using (\ref{schrodapprox}) we obtain for any $\psi\in\H_+$
\bea
\label{diff_app}
\partial_r \left[U_n(t,r) U_m(r,s)\psi\right]=i \; U_n(t,r) [A_n(r)-A_m(r)] U_m(r,s)\psi\,,
\eea
for $r\neq \frac{j T}{n}, r\neq \frac{j T}{m}$ with $j=1,\cdots,\max(n,m)$. Integrating (\ref{diff_app}) we get the identity
$$
U_m(t,s)\psi-U_n(t,s)\psi=i\,\int_s^t U_n(t,r) \, [A_n(r)-A_m(r)] \,U_m(r,s) \psi\;dr\,.
$$
Now (\ref{kat-est}) yields
\bea
\label{propag-est}
||U_m(t,s)-U_n(t,s)||_{\L(\H_+,\H_-)}\leq M^2  |t-s|e^{2c |t-s|} \;\sup_{r\in I}||A_m(r)-A_n(r)||_{\L(\H_+,\H_-)} \,.
\eea
Therefore, for any $t,s\in I$, the sequence $U_n(t,s)$ converges in norm to a bounded linear operator $U(t,s)\in\L(\H_+,\H_-)$.
Since $U_n(t,s)$ are norm bounded operators on $\H_-$ uniformly in $n$, it follows by (\ref{kat-est}) that they converge strongly to
an operator in $\L(\H_-)$ continuously extending  $U(t,s)$. Moreover, this strong convergence yields
\bean
\lim_{n\to\infty} (\phi,U_n(t,s)\psi)_{\H}=(\phi,U(t,s)\psi)_{\H} \quad \forall \psi\in\H_+, \forall \phi\in\H_+\,.
\eean
where $(.,.)_\H$ is the continuous extension of the inner product of $\H$ to the rigged Hilbert spaces $\H_\pm$.
Thus, using (\ref{kat-est}), we obtain
\bean
\left|(\phi,U(t,s)\psi)_{\H}\right| \leq M e^{c|t-s|} ||\phi||_{\H_-} \;||\psi||_{\H_+}\,.
\eean
Hence, it is easy to see by (\ref{rigged2}) that
\bean
||U(t,s)||_{\L(\H_+)}\leq M e^{c|t-s|} \,.
\eean
A similar argument yields
\bea
\label{unitun}
||U(t,s)||_{\L(\H)}\leq \,1\,.
\eea
Now, since $U_n(t,s)$ satisfy part (b) of Definition \ref{unit-propag}, we easily conclude that
\bea
\label{2gplaw}
U(t,t)=1, \quad U(t,r) U(r,s)=U(t,s),\quad t,s,r\in I,
\eea
by strong convergence in $\L(\H_{-})$. Furthermore, combining (\ref{unitun}) and (\ref{2gplaw}) we show
the unitarity of $U(t,s)$ on $\H$.  Thus, we have proved that $U(t,s)$ satisfy (a)-(b) of Definition \ref{unit-propag}.

For any $\psi\in\H_+$, the continuity of the map $I\ni t\mapsto U_n(t,s)\psi\in\H_-$ follows from the definition of $U_n(t,s)$.
Now, we prove
\bean
\lim_{t\to s} (\phi, U(t,s)\psi)_{\H} =(\phi,\psi)_{\H} \quad \forall \psi\in \H_+, \forall\phi\in\H_-\,,
\eean
by applying an $\epsilon/3$ argument when writing
\bean
\left| (\phi,U(t,s)\psi)_{\H}-(\phi,\psi)_{\H}\right|&\leq &
||\phi-\phi_\kappa||_{\H_-} ||U(t,s)\psi||_{\H+}+ \left|(\phi_\kappa, [U(t,s)-U_n(t,s)]\psi)_\H\right| \\
&&+
\left|(\phi_\kappa, [U_n(t,s)-1] \psi)_\H\right|+ ||\phi-\phi_\kappa||_{\H_-} \||\psi||_{\H_+}\,,
\eean
where $\phi_\kappa \to\phi$ in $\H_-$.  Therefore, by the duality $(\H_+)'\simeq\H_-$, we get the weak limit
$$
w-\lim_{t\to s} U(t,s)=1\,,
$$
in $\L(\H_+)$. Now, observe that when $t\to s$  we can show by (\ref{kat-est}) that
$$
\limsup_{t\to s} \;||U(t,s)\psi||_{\H_+} \leq ||\psi||_{\H_+}\,.
$$
So, we conclude that
\bean
\limsup_{t\to s}\;||U(t,s)\psi-\psi||^2_{\H_+}\leq \limsup_{t\to s} \left(
||\psi||_{\H_+}^2+||U(t,s)\psi||^2_{\H_+} -2 {\rm Re}(\psi,U(t,s)\psi)_{\H_+}\right)=0\,.
\eean
This gives the continuity of $I\ni t\mapsto U(t,s)\psi\in\H_+$ since  we have in $\H_+$
$$
s-\lim_{t \to r} U(t,s)=s-\lim_{t\to r}U(t,r)U(r,s)=U(r,s).
$$
Now, we have for $\psi\in\H_+$ as identity in $\H_-$
\bea
\label{diff-As}
e^{-i\tau A(s)}\psi=\psi-i A(s)\,\int_0^\tau e^{-i r A(s)} \psi \,dr\,,
\eea
since this holds first for $\psi\in\D(A(s))\cap\H_+$ and then extends by density of $\D(A(s))\cap\H_+$ in $\H_+$. 
By (\ref{diff-As}) we have 
\bean
||\frac{e^{-i\tau A(s)}\psi-\psi}{\tau}+iA(s)\psi||_{\H_-}\leq \frac{1}{\tau} ||A(s)||_{\L(\H_+,\H_-)} 
\, \left|\int_0^\tau ||e^{-i r A(s)} \psi-\psi||_{\H_+} \,dr\right|
\eean  
and hence using assumption (ii), we show the differentiability of $\tau\mapsto e^{-i\tau A(s)}\psi$ for $\psi\in\H_+$. 
By differentiating $e^{-i (t-r) A(s)} U_m(r,s)\psi$ with $\psi\in\H_+$ and then integrating w.r.t. $r$, we get
$$
U_m(t,s)\psi-e^{-i (t-s) A(s)} \psi=i\, \int_s^t e^{-i (t-r) A(s)} [A(s) -A_m(r)] U_m(r,s)\psi \,dr\,.
$$
Letting $m\to\infty$ in the latter identity and estimating as in (\ref{propag-est}), one obtains
$$
||U(t,s)\psi-e^{-i (t-s) A(s)} \psi||_{\H_-}\leq M^2 e^{2c |t-s|}\left| \int_s^t || [A(s) -A(r)||_{\L(\H_+,\H_-)}  \,dr\right|
\,||\psi||_{\H_+}.
$$
Using the fact that
$$
\lim_{t\to s} \frac{1}{|t-s|} \int_s^t ||A(s)-A(r)||_{\L(\H_+,\H_-)} dr=0 \quad \mbox{ and } \quad
\lim_{t\to s} \frac{e^{-i (t-s) A(s)} \psi-\psi}{t-s}=-i A(s) \psi 
$$
it holds that
$$
\lim_{t\to s} \left\|\frac{U(t,s) \psi-\psi}{t-s}+i A(s) \psi\right\|_{\H_-}=0.
$$
Thus, we obtain with the help of (\ref{2gplaw})
$$
i\partial_s U(s,r)\psi= \lim_{t\to s} \frac{U(t,s) U(s,r)\psi-U(s,r)\psi}{t-s}=A(s) U(s,r)\psi,
$$
for any $\psi\in\H_+$ and any $r,s\in I$. Hence we have proved the existence of a unitary propagator $U(t,s)$ 
for the non-autonomous Cauchy problem (\ref{abs_schrod}).

\bigskip
\noindent
\underline{\bf Uniqueness:} \\
Suppose that $V(t,s)$ is a unitary propagator for (\ref{abs_schrod}). By differentiating $ U_n(t,r)V(r,s)\psi$, $\psi\in\H_+$ 
with respect to  $r$ we get
$$
V(t,s)\psi-U_n(t,s)\psi=i\int_s^t   U_n(t,r) [A_n(r)-A(r)] V(r,s)\psi.
$$
Using a similar estimate as (\ref{propag-est}) we obtain
$$
\|V(t,s)\psi-U_{n}(t,s)\psi\|_{\mathcal{H}_{+}}\leq Me^{c|t-s|}\sup_{r\in\left[s,t\right]}\left\Vert V(r,s)\right\Vert _{\mathcal{L}(\mathcal{H}_{+})}\left|\int_{s}^{t}||A(r)-A_{n}(r)||_{\mathcal{L}(\mathcal{H}_{+},\mathcal{H}_{+})}dr\right|\,||\psi||_{\mathcal{H}_{+}}
$$
and since the r.h.s. vanishes when $n\to\infty$ we conclude that $V(t,s)=U(t,s)$. \\
Finally, the uniform boundedness principle, equivalence of norms $||.||_t,||.||_{\H_+}$ and the inequality (\ref{gen-propag-est}) give us 
the claimed estimate,
\bean
\forall \psi\in\H_+,\quad ||U(t,s)\psi||_t\leq \liminf_{n\to\infty}\, ||U_n(t,s)\psi||_t\leq e^{2c |t-s|} ||\psi||_s\,.
\eean
\hfill\cqfd

\begin{remark}
It also follows that $(t,s)\mapsto U(t,s)\in\L(\H_+)$ is jointly strongly continuous.
\end{remark}

\bigskip
In the following  we provide a more effective formulation of the
above result (Theorem \ref{thm_schrod}) which appears as a time-dependent version of the Nelson commutator theorem (see, {\it e.g.}, \cite{N},
\cite{RS} and Theorem \ref{nelson}).

\bigskip

We associate to each family of self-adjoint  operators
$\{S(t)_{t\in I},S\}$ on $\H$ such that $S\geq 1$, $S(t)\geq 1$ and $\D(S(t)^{1/2})=\D(S^{1/2})$ for any $t\in I$,
a Hilbert rigging $\H_{\pm 1}$ defined as the completion of $\D(S^{\pm 1/2})$ with respect to the inner product
\bea
\label{inner_scale}
\la \psi,\phi\ra_{\H_{\pm 1}} =\la S^{\pm 1/2} \psi,S^{\pm 1/2}\phi\ra_\H.
\eea

\begin{cor}
\label{abs_schrod_cor}
Let $I\subset \rz$ be a closed interval and let $\{S(t)_{t\in I},S\}$  be a family of self-adjoint operators on a Hilbert space
$\H$ such that:
\begin{itemize}
 \item $S\geq 1$ and $ S(t)\geq 1$, $\forall t\in I$,
 \item $\D(S(t)^{1/2})=\D(S^{1/2})$,  $\forall t\in I$, and consider the associated Hilbert rigging $\H_{\pm 1}$ given by (\ref{inner_scale}).
\end{itemize}
Let  $\{A(t)\}_{t\in I}$ be a family of symmetric bounded operators in $\L(\H_{+1},\H_{-1})$  satisfying:
\begin{itemize}
 \item $t\in I\mapsto A(t)\in \L(\H_{+1},\H_{-1})$ is norm continuous.
\end{itemize}
Assume that there exists a continuous function $f:I\to\rz_{+}$ such that for any $t\in I$, we have:\\
(i) for any $\psi\in \D(S(t)^{1/2})$,
\bean
|\partial_t\la \psi,  S(t) \psi\ra |\leq f(t) \;||S(t)^{1/2}\psi||^2\,;
\eean
(ii) for any $\Phi,\Psi\in\D(S(t)^{3/2})$,
\bean
\left|\la S(t)\Psi, A(t)\Phi\ra-\la A(t)\Psi,S(t)\Phi\ra \right|\leq f(t) \,||S(t)^{1/2}\Psi||\;||S(t)^{1/2}\Phi||.
\eean
Then the non-autonomous Cauchy problem (\ref{abs_schrod}) admits a unique unitary propagator $U(t,s)$.
Moreover, we have
\bean
||S(t)^{1/2}U(t,s)\psi||\leq e^{2 \;|\int_s^t f(\tau) d\tau| } \;\;||S(s)^{1/2}\psi||\,.
\eean
In addition, if we have $c_1,c_2>0$ such that $\ds c_1 S\leq S(t)\leq c_2 S$ for $ t\in I$, then there exists $c>0$
such that
\bea
\label{est-nelst}
||U(t,s)||_{\L(\H_{+1})}\leq c\;e^{2|\int_s^t f(\tau)\d\tau|} \;, \quad \forall  t\in I\,.
\eea
\end{cor}
\proof
First observe  that the operator $A(t)$ satisfies the hypothesis of Nelson's commutator theorem (Theorem \ref{nelson}) for any
$t\in I$. Hence, we conclude that $A(t)$ is essentially self-adjoint  on $\D(S(t)^{3/2})$ 
which is dense in $\H_{+1}$. We keep the same notation for its closure.
Moreover, the unitary group $e^{i\tau A(t)}$ preserves $\H_{+1}$ and we have the estimate
\bea
\label{asumpnels2}
||S(t)^{1/2} e^{i\tau A(t)}\psi||_{\H}\leq e^{f(t) |\tau|} \;||\psi||_{\H}\,.
\eea
Now, observe that $t\mapsto e^{-it A(s)}\psi\in\H_{+1}$ is weakly continuous for any $\psi\in\H_+$. This 
holds using a $\eta/3$-argument with the help of the estimate
\bean
\left|\la f,(e^{-it A(s)}-1)\psi\ra\right|\leq (1+e^{c (|t|+1)}) \, 
||f-f_\kappa||_{\H_{-1}} \,||\psi||_{\H_{+1}}+\left|\la (e^{it A(s)}-1)f_\kappa,\psi\ra\right|
\eean 
where $f_\kappa\in\H$ is a sequence convergent to $f$ in $\H_{-1}$ and $t$ is near $0$. Since strong and weak 
continuity of the group of bounded operators $e^{-it A(s)}$ in $\L(\H_{+1})$ are equivalent, we conclude 
that assumption (ii) of Theorem  \ref{thm_schrod} holds true.

\noindent
By assumption (ii), we also have
\bean
\left|\frac{d}{dt} ||S(t)^{1/2} \psi||^{2} \right|\leq  f(t) ||S(t)^{1/2} \psi||^2\,.
\eean
Hence, by Gronwall's inequality we have
\bea
\label{asumpnels1}
\ds||S(t)^{1/2}\psi||^2 \leq e^{|\int_s^t f(\tau) d\tau|} ||S(s)^{1/2}\psi||^2, \quad\forall t,s\in I.
\eea
Now, we  use Theorem \ref{thm_schrod} with the Hilbert rigging
$$
\H_+=\H_{+1}\subset \H\subset \H_-=\H_{-1}
$$
and the family of equivalent norms on $\H_+$ given by
$$
||\psi||_t:=||S(t)^{1/2}\psi||_{\H}.
$$
Indeed, assumptions (i)-(iii) of Theorem  \ref{thm_schrod} are satisfied in any compact subinterval of $I$ with the help of
(\ref{asumpnels1})-(\ref{asumpnels2}). Therefore, we obtain  existence and uniqueness
of a unitary propagator $U(t,s)$ of the Cauchy problem (\ref{abs_schrod}) in the whole interval $I$ with the following estimate 
\bean
||U(t,s)\psi||_t\leq \ds e^{2|t-s|\; \max_{\tau\in \Delta(t,s)}f(\tau) } \;\;||\psi||_s\,,
\eean
for any $t,s\in I$ and where $\Delta(t,s)$ stands for the interval of extremities $t$, $s$.

Using the multiplication law of the propagator,  we obtain
for any partition $(t_0,\cdots,t_n)$ of the interval $\Delta(t,s)$ the inequality
\bean
||U(t,s)\psi||_t\leq \prod_{j=0}^{n-1} e^{2 \;\frac{|t-s|}{n} \max_{\tau\in {\Delta_j}}f(\tau) } \;\;||\psi||_s\,,
\eean
where $\Delta_j$ are the subintervals $[t_j,t_{j+1}]$. Since $f$ is continuous, by letting $n\to\infty$, we get
\bean
||U(t,s)\psi||_t\leq e^{2 \;|\int_s^t f(\tau) d\tau| } \;\;||\psi||_s\,.
\eean
Finally, the assumption  $\ds c_1 S\leq S(t)\leq c_2 S$ for $ t\in I$, allows to involve the norm $||.||_{\H_{+1}}$. Thus we have
\bean
||U(t,s)\psi||_{\H_{+1}}\leq \frac{1}{\sqrt{c_1}} ||U(t,s)\psi||_t\leq \frac{1}{\sqrt{c_1}}
e^{2 \;|\int_s^t f(\tau) d\tau| } \;\;||\psi||_s \leq \sqrt{ \frac{c_2}{c_1}}  \;e^{2 \;|\int_s^t f(\tau) d\tau| } \;\;||\psi||_{\H_{+1}}\,.
\eean
\hfill\cqfd

\bigskip
\noindent\textbf{Acknowledgements:}
The authors would like to thank Francis Nier for very useful and stimulating discussions.

\bibliographystyle{empty}

\begin{thebibliography}{999}

\bibitem[ABGT]{ABGT} R.~Adami, C.~Bardos, F.~Golse, A.~Teta.
\newblock Towards a rigorous derivation of the cubic NLSE in dimension one.
\newblock Asymptot.~Anal. 40, 93--108  (2004)

\bibitem[AGT]{AGT} R.~Adami, F.~Golse, A.~Teta.
\newblock Rigorous Derivation of the Cubic NLS in Dimension One.
\newblock  J.~Stat.~Physics 127, 1193--1220 (2007)

\bibitem[AmNi1]{AmNi1} Z.~Ammari, F.~Nier.
\newblock Mean field limit for bosons and infinite dimensional phase-space analysis.
\newblock Ann.~Henri Poincar\'e 9 (2008), 1503--1574

\bibitem[AmNi2]{AmNi2} Z.~Ammari, F.~Nier.
\newblock Mean field limit for bosons and propagation of Wigner measures.
\newblock J.~Math.~Phys. 50 (2009)

\bibitem[BGM]{BGM} C.~Bardos, F.~Golse, N.~Mauser.
\newblock Weak coupling limit of the n-particle Schr\"odinger equation.
\newblock Methods Appl. Anal. 7, 275--293 (2000)

\bibitem[BEGMY]{BEGMY} C.~Bardos, L.~Erd\"{o}s, F.~Golse, N.~Mauser, H-T.~Yau.
\newblock Derivation of the Schr\"odinger-Poisson equation from the quantum N-body problem.
\newblock C.R. Math. Acad. Sci. Paris 334, 515--520 (2002)

\bibitem[CRR]{CRR} M.~Combescure, J.~Ralston, D.~Robert.
\newblock  A proof of the Gutzwiller semiclassical trace formula using coherent states decomposition.
\newblock Comm. Math. Phys. 202 (1999), no. 2, 463-480.

\bibitem[DA]{DA} G.~F.~Dell'Antonio.
\newblock On the limits of sequences of normal states.
\newblock  Comm. Pure Appl. Math., 20 (1967),  413--429.

\bibitem[ElSc]{ElSc} A. Elgart, B. Schlein.
\newblock Mean field dynamics of boson stars
\newblock Comm. Pure and Appl. Math. Vol. 60, (2007) 500--545

\bibitem[EY]{EY} L.~Erd{\"o}s, H.T.~Yau.
\newblock Derivation of the nonlinear Schr\"odinger equation from a many body coulomb system.
\newblock Adv. Theor. Math. Phys. 5, 1169--2005 (2001)

\bibitem[ESY]{ESY} L.~Erd{\"o}s, B.~Schlein, H.T.~Yau.
\newblock Derivation of the cubic non-linear Schr{\"o}dinger equation from quantum dynamics of many-body systems.
\newblock Invent. Math. 167 (2007), no. 3, 515--614.

\bibitem[FGS]{FGS} J.~Fr\"{o}hlich, S.~Graffi, S.~Schwarz.
\newblock Mean-field- and classical limit of many-body
Schr\"{o}dinger dynamics for bosons.
\newblock Comm. Math. Phys.  271, No. 3 (2007), 681--697.

\bibitem[FKP]{FKP} J.~Fr\"ohlich, A.~Knowles, A.~Pizzo.
\newblock Atomism and quantization.
\newblock  J. Phys. A 40, no. 12 (2007), 3033--3045.

\bibitem[FKS]{FKS} J.~Fr\"{o}hlich, A.~Knowles, S.~Schwarz.
\newblock On the Mean-field limit of bosons with Coulomb two-body interaction
\newblock http://arxiv.org/abs/0805.4299

\bibitem[FL]{FL} W.~Faris, R.~Lavine.
\newblock Commutators and selfadjointness of Hamiltonian operators.
\newblock   Comm. Math. Phys. 35 (1974), 39--48

\bibitem[GiVe1]{GiVe1}J.~Ginibre, G.~Velo.
\newblock The classical field limit of scattering theory for nonrelativistic
many-boson systems. I.
\newblock Comm. Math. Phys. 66 (1979), 37--76

\bibitem[GiVe2]{GiVe2}J.~Ginibre, G.~Velo.
\newblock The classical field limit of scattering theory for nonrelativistic
many-boson systems. II.
\newblock Comm. Math. Phys. 68, (1979), 45--68

\bibitem[GiVe3]{GiVe3}J.~Ginibre, G.~Velo.
\newblock On a class of nonlinear Schr\"odinger equations. I. The Cauchy problem. General case.
\newblock J. Functional Anal. 32, (1979), 1--32

\bibitem[Gol]{Gol}F.~Golse.
\newblock The mean-field limit for the dynamics of large particle systems.
\newblock Journ\'ees \'equations aux d\'eriv\'ees partielles (2003), Art. No. 9, 47 p.

\bibitem[Go]{Got} A.~D.~Gottlieb.
\newblock Propagation of chaos in classical and quantum kinetics.
\newblock Stochastic analysis and mathematical physics II,  135--146, Trends Math., Birkh\"auser, Basel, 2003.

\bibitem[Hep]{Hep}K.~Hepp.
\newblock The classical limit for quantum  mechanical correlation functions.
\newblock  Comm.~Math.~Phys. 35 (1974), 265--277

\bibitem[Ka]{Ka} T.~Kato.
\newblock Linear evolution equations of ``hyperbolic'' type.
\newblock J.~Fac.~Sci.~Univ.~Tokyo Sect.~I  17,  (1970), 241--258

\bibitem[Ki]{Ki} J.~Kisynski.
\newblock Sur les op\'erateurs de Green des probl\`emes de Cauchy abstraits.
\newblock Studia Math.~23 (1963/1964), 285--328

\bibitem[N]{N} E.~Nelson.
\newblock Time-Ordered Operator Products of Sharp-Time Quadratic Forms.
\newblock  J.~Functional Anal.~11, 211--219 (1972)

\bibitem[MS]{MaSh} V.P.~Maslov, O.Y. Shvedov.
\newblock The chaos conservation problem in quantum physics.
\newblock Russian J.~Math.~Phys.  4  (1996),  no. 2, 173--216.

\bibitem[RS]{RS} M.~Reed, B.~Simon.
\newblock Methods of Modern Mathematical Physics, vol.~II
\newblock Academic Press (1976).

\bibitem[RoSch]{RodSch} I.~Rodnianski, B.~Schlein.
\newblock Quantum Fluctuations and Rate of Convergence towards Mean Field Dynamics.
\newblock http://arxiv.org/abs/0711.3087

\bibitem[Si]{Si} B.~Simon.
\newblock Quantum Mechanics for Hamiltonians Defined as Quadratic Forms.
\newblock Princeton University Press, Princeton, N.~J., 1971

\bibitem[Spo]{Spo}H.~Spohn.
\newblock Kinetic equations from Hamiltonian dynamics.
\newblock Rev. Mod. Phys. 52, No. 3 (1980), 569--615

\bibitem[T]{T} Y.~Tsutsumi.
\newblock Global strong solutions for nonlinear Schr\"odinger equations.
\newblock  Nonlinear Anal.  11, No. 10  (1987), 1143--1154
\end{thebibliography}

\end{document}